\documentclass[useAMS,usenatbib, usegraphicx]{mn2e}
\pdfoutput=1
\usepackage{fancyhdr}
\usepackage{lscape}
\usepackage{hyperref}
\usepackage{wasysym}
\usepackage{longtable}
\usepackage{aas_macros}
\usepackage{txfonts}
\usepackage{times}
\usepackage{graphicx}
\usepackage{color}
\usepackage{ulem}

\begin{document}
\title[Spur, blob, wiggle, and gaps in GD-1]{A closer look at the spur, blob, wiggle, and gaps in GD-1}
\author[T.J.L. de Boer et al.]{T.J.L. de Boer$^{1,2}$\thanks{E-mail:
tdeboer@hawaii.edu}, D. Erkal$^{2}$ and M. Gieles$^{2,3,4}$\\
$^1$ Institute for Astronomy, University of Hawai`i, 2680 Woodlawn Drive, Honolulu, HI 96822, USA\\
$^2$ Department of Physics, University of Surrey, Guildford, GU2 7XH, UK\\
$^3$ Institut de Ci\`{e}ncies del Cosmos (ICCUB-IEEC), Universitat de Barcelona, Mart\'{i} i Franqu\`{e}s 1, 08028 Barcelona, Spain\\
$^4$ ICREA, Pg. Lluis Companys 23, 08010 Barcelona, Spain.\\
}
\date{Received ...; accepted ...}

\pagerange{\pageref{firstpage}--\pageref{lastpage}} \pubyear{2017}

\maketitle

\begin{abstract}
The GD-1 stream is one of the longest and coldest stellar streams discovered to date, and one of the best objects for constraining the dark matter properties of the Milky Way. Using data from {\it Gaia} DR2 we study the proper motions, distance, morphology and density of the stream to uncover small scale perturbations. The proper motion cleaned data shows a clear distance gradient across the stream, ranging from 7 to 12 kpc. However, unlike earlier studies that found a continuous gradient, we uncover a distance minimum at $\varphi_{1}\approx$-50 deg, after which the distance increases again. We can reliably trace the stream between -85$<\varphi_{1}<$15 deg, showing an even further extent to GD-1 beyond the earlier extension of \citet{Price-Whelan18a}.
We constrain the stream track and density using a Boolean matched filter approach and find three large under densities and find significant residuals in the stream track lining up with these gaps. In particular, a gap is visible at $\varphi_{1}$=-3 deg, surrounded by a clear sinusoidal wiggle. We argue that this wiggle is due to a perturbation since it has the wrong orientation to come from a progenitor. We compute a total initial stellar mass of the stream segment of 1.58$\pm$0.07$\times$10$^{4}$ M$_{\odot}$. With the extended view of the spur in this work, we argue that the spur may be unrelated to the adjacent gap in the stream. Finally, we show that an interaction with the Sagittarius dwarf can create features similar to the spur.
\end{abstract}

\begin{keywords}
Galaxy: structure -- Galaxy: fundamental parameters --- Stars: C-M diagrams --- Galaxy: halo --- Galaxy: kinematics and dynamics
\end{keywords}

\label{firstpage}

\section{Introduction}
\label{introduction}
Stellar streams are rivers of stars formed from the disruption of stellar systems within a larger host galaxy due to the effects of the host gravitational field. Their location and appearance is uniquely governed by the host's gravitational field, which makes them one of the best tracers for measuring its properties. In general, the mean track and phase-space density of the streams encode information about the underlying unseen distribution of mass on galactic scales \citep[e.g.,][]{Johnston99}. However, for systems with a suitably small initial phase-space volume (such as globular clusters), the stellar streams can also retain imprints on from encounters with other perturbers, such as small-scale dark matter (DM) subhaloes \citep[e.g.][]{Johnston02,Ibata02,Carlberg09,Yoon11,Erkal16,Sanders16,Bovy17}. Therefore, stellar streams provide not just dramatic confirmation of the hierarchical galaxy formation scenario, but can also be directly used to quantify the properties of the (unseen) matter distribution \citep{Lynden-Bell95,Ibata02,Johnston02}.

Although initially rare, the gallery of stellar streams has expanded greatly during the last two decades with the availability of large scale, homogeneous sky surveys. Discoveries of streams have been made in all the large surveys such as SDSS, 2MASS, Pan-STARRS, ATLAS, and DES \citep{Skrutskie06,Ahn14,Shanks15,Chambers16}, after applying photometric selection methods specifically designed to find streams \citep{Grillmair16}. \textit{Gaia} has provided exquisite data which has been used to discover a wealth of streams \citep[e.g.][]{malhan_2018,ibata_phlegethon,ibata_2019}. At present, more than 60 streams have been found in the Milky Way (MW) alone, with wildly different morphologies ranging from multiple wraps (Sagittarius) to the very short (Ophiuchus) and anything in between \citep{Belokurov06,Bernard14}.

Since the initial survey finds, individual streams have been used to place constraints on the mass distribution of the MW halo, confirming they are a powerful tool for probing the unseen MW components \citep[e.g.,][]{Koposov10,Law10a,Gibbons14,Bowden15,Kuepper15,Bovy16}. Furthermore, in recent years detailed stream morphology has been modelled to probe for the existence of dark matter~(DM) subhalos \citep[e.g.][]{Carlberg09,Yoon11,Carlberg12,Erkal15}, which are not observable by conventional methods \citep{Ikeuchi86,Rees86}. Indeed, tentative evidence for a disturbance by low mass subhaloes has been found in the Pal 5 stream \citep{Bovy17,Erkal17}.

One of the difficulties in uncovering the origin of details in the stream morphology is related to the uniqueness of the signature left by an encounter with a dark subhalo \citep{Yoon11,Carlberg12,Erkal15,Sanders16}. For some streams (including Pal 5), interaction with a giant molecular cloud, the MW bar, or spiral arms \citep{Amorisco16, Erkal17, Pearson17,banik_bovy_2019} can result in similar signatures as those left by a close subhalo passage. This complicates the otherwise straightforward interpretation of stream features, and has led to the search for a suitable stream in which the other explanations are excluded or unlikely due to their distance or orbit.

The GD-1 stream is one of the most promising streams for the study of DM on small scales, given its morphology and orbit. When first discovered in SDSS, it was found as a 63 degree long, very thin stream with a width of only 0.2 deg \citep{Grillmair061}. This means it has a small initial phase-space volume, which allows us to more easily distinguish velocity kicks induced by dark substructures from the secular dynamics of the progenitor. Furthermore, the GD-1 stream is moving in a retrograde sense with respect to the disk, which means interactions with MW disk sub-strucure are less effective in creating stream features \citep{Koposov10,Bowden15}. The orbit of the stream (with a perigalacticon of 14 kpc and apogalacticon of 26-29 kpc) is such that interactions with giant molecular clouds or the MW bar are also unlikely. Finally, the stream is on a retrograde orbit making such interactions negligible \citep[e.g.][]{Amorisco16}, making GD-1 an ideal candidate for the study of gaps induced by dark subhaloes.

\begin{figure}
\centering
\includegraphics[angle=0, width=0.495\textwidth]{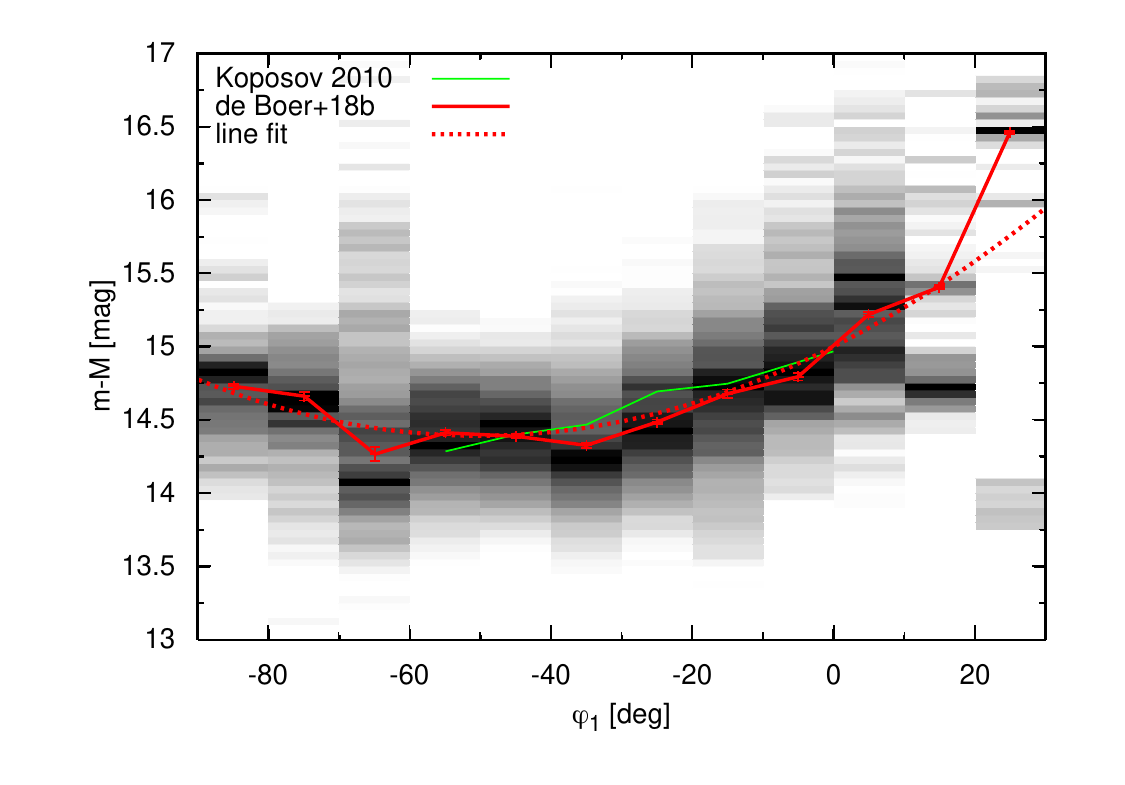}
\caption{Distance determination for GD-1 as a function of $\varphi_{1}$, showing the column normalised density of stars in bins of distance modulus. The green line shows literature distances from \citet{Koposov10}. The red points indicate the best-fit Gaussian distance peak to each bin of $\varphi_{1}$. The dotted red line shows a best fit second order polynomial to the distance, as described by $f(\varphi_{1})=15.001+2.421\times10^{-2}\varphi_{1}+2.410\times10^{-4}\varphi_{1}^{2}$. \label{GD1_distfit}}
\end{figure}

Several previous studies of GD-1 have revealed that it displays a complex morphology, despite its initial simple appearance. Using SDSS data, \cite{Carlberg13} found a large number of gaps in GD-1. A study of deep photometric data from the Canada-France-Hawaii-Telescope (CFHT) by \citet{deBoer2018a} showed that there are at least two clear gaps where the density is consistent with zero given the uncertainties, and the stream track shows deviations that could be due to interactions. Earlier this year, data from the recently released {\it Gaia} DR2 was used to identify stream members with much greater accuracy, enabling the discovery of a stream extension (the total length is now in excess of 90 degrees) and two associated but separate stream structure, dubbed the spur and the blob \citep{Price-Whelan18a}. The connection between the spur feature and the main stream (in a location where a stream gap was found) is a prime candidate for DM interaction, as recently modelled by \citet{Bonaca18}. Independently, \citet{Webb18} also found that the GD-1 stream extends from $-90<\varphi_{1}<10$ degrees, which makes it even longer than the \citet{Price-Whelan18a} detection, and measure the track and density. 

In this work, we will revisit the investigation of GD-1 using {\it Gaia} DR2 data, and employ a more sophisticated data filtering using a matched filter approach \citep{Rockosi02,deBoer2018a}. The aim of the present work is to extract the maximum of information from the {\it Gaia} DR2  and search for small scale variations in the stream track. We will constrain the distance to the stream across (and beyond) it current length and fit the proper motion signal as a function of stream angle.  The evolution of the stream centroid of the sky will be constrained, which will yield wiggles and gaps, all of which are indicative of perturbations due to tidal encounters \citep{Carlberg12,Erkal15,Erkal15b}. We will also study the stream density profile and determine the total stellar mass encompassed by GD-1. This data will be used in two companion papers (Banik et al. in prep a, Banik et al. in prep b) to explore the dark matter substructure in the Milky Way and place a lower bound on mass of the dark matter particle.

This paper is organised as follows: in Section~\ref{data} we discuss the data used, followed by the determination of the distance to GD-1 in Section~\ref{distance}. Section~\ref{proper_motions} then discusses the PM signal of the stream and how this was fit, with the determination of the stream track locus in Section~\ref{stream_track}. The density and mass of GD-1 is subsequently determined in Section~\ref{stream_density}, and the residuals in location and PMs is discussed in Section~\ref{residuals}. In Section~\ref{sec:discussion} we fit this data with an orbit, explore how well the velocities of GD-1 are aligned with the stream, show that an interaction with Sagittarius can create features similar to the spur, and discuss the morphology of features in GD-1. Finally, Section~\ref{conclusions} discusses the results and their implications.

\section{Data}\label{data}
To investigate the GD-1 stream, we make use of data from the {\it Gaia} mission~\citep{GAIAmain1,GAIAmain2,Lindegren18}, which contains proper motion (PM) measurements for $\approx$1.6 billion sources covering the full sky. The astrometry of {\it Gaia} are excellent, but the spectro-photometric colours suffer from systematics in crowded regions. Therefore, we combine the {\it Gaia} astrometry with the wide-field photometry of the Pan-STARRS survey, data release 1~\citep{Chambers16}. Following this, the photometry of Pan-STARRS is corrected for extinction using dust maps from~\citet{Schlegel98} with coefficients from \citet{Schlafly11}, on a star by star basis. 

We adopt the stream-aligned coordinate scheme of \citet{Koposov10}, converting ra,dec into $\varphi_{1},\varphi_{2}$ representing along and perpendicular to the stream respectively. Furthermore, the {\it Gaia} PMs $\mu_{\mathrm{ra}}$, $\mu_{\mathrm{dec}}$ are converted to $\mu_{\mathrm{\varphi_{1}}}$, $\mu_{\mathrm{\varphi_{2}}}$ representing the PM in the stream-aligned frame.

\subsection{Distance to the stream}\label{distance}
A crucial step in extracting the best sample of stream stars is determining an accurate distance to the individual parts of the stream. The distance to GD-1 has been determined using deep CFHT data for a small portion of the stream \citep{deBoer2018a} and for a longer stretch of stream from SDSS CMDs \citep{Koposov10}. Across the range of $\varphi_{1}$ for which distances could be derived by SDSS data ($-55<\varphi_{1}<0$ deg) GD-1 displays a linear distance variation between 7.5 and 10 kpc. 

To investigate the distances beyond the range probed by SDSS, we compare the observed magnitude of main sequence stars to those of a reference isochrone with [Fe/H]=-1.9, age=12 Gyr from the Padova library \citep{Marigo17}, as queried from \url{http://stev.oapd.inaf.it/cmd}. To select GD-1 stars, we apply rough cuts in PM and consider only stars within 2 degrees of the stream track based on results from \citet{Price-Whelan18a}. We then sample small bins of distance modulus and sum up the number of stars consistent with the reference isochrone within twice the photometric errors. To bring out the signal of the GD-1 stream, we correct for the effects of MW contamination by performing the same analysis for stars located between 2-4 degrees away from the track on either side, and subtracting those. 

Figure~\ref{GD1_distfit} shows the density of stars in bins of distance modulus, as a function of $\varphi_{1}$. The red points indicate the best-fit distance from a Gaussian fit to each bin of $\varphi_{1}$. For reference, the distance computed by \citet{Koposov10} are shown as the green line. It is apparent that the recovered distance trend is not linearly varying in the {\it Gaia} data, but instead recedes to larger distances beyond the $\varphi_{1}$ range sampled by SDSS. This can have important implications for the orbit of the stream, and the number of interactions with subhaloes at smaller distances. Interestingly, the recovered location of the distance minimum of $\varphi_{1}\approx$-40 degrees corresponds very well to the predicted minimum of from the dynamical model of GD-1 by \citet{Bovy16}.
\begin{table*}
\caption[]{Distance modulus, heliocentric distance and PMs (along with Gaussian widths) of the GD-1 stream from {\it Gaia} DR2.}
\begin{center}
\begin{tabular}{ccccccc}
\hline\hline
$\varphi_{1}$ &  m-M & distance & $\mu_{\mathrm{\varphi_{1}}}$ & $\mu_{\mathrm{\varphi_{2}}}$ & $\sigma_{\mu_{\mathrm{\varphi_{1}}}}$ & $\sigma_{\mu_{\mathrm{\varphi_{2}}}}$ \\
{[deg]} & [mag] & [kpc] & [mas/yr] & [mas/yr] & [mas/yr] & [mas/yr]  \\
\hline
-85 & 14.73$\pm$0.02 &  8.84$\pm$0.07 &  -9.66$\pm$0.28 & -3.29$\pm$0.26 & 0.89$\pm$0.31 & 0.62$\pm$0.21 \\ 
-75 & 14.66$\pm$0.03 &  8.57$\pm$0.12 & -10.90$\pm$0.15 & -3.65$\pm$0.15 & 0.40$\pm$0.11 & 0.63$\pm$0.16 \\
-65 & 14.27$\pm$0.05 &  7.14$\pm$0.16 & -11.89$\pm$0.34 & -4.05$\pm$0.33 & 1.22$\pm$0.37 & 0.93$\pm$0.28 \\
-55 & 14.41$\pm$0.02 &  7.63$\pm$0.07 & -13.25$\pm$0.06 & -3.49$\pm$0.09 & 0.51$\pm$0.05 & 0.82$\pm$0.08 \\
-45 & 14.39$\pm$0.02 &  7.55$\pm$0.05 & -13.35$\pm$0.04 & -3.38$\pm$0.09 & 0.34$\pm$0.03 & 0.82$\pm$0.07 \\
-35 & 14.33$\pm$0.02 &  7.34$\pm$0.06 & -13.05$\pm$0.06 & -3.15$\pm$0.08 & 0.33$\pm$0.04 & 0.64$\pm$0.08 \\
-25 & 14.49$\pm$0.02 &  7.90$\pm$0.06 & -12.35$\pm$0.06 & -2.80$\pm$0.07 & 0.44$\pm$0.05 & 0.59$\pm$0.07 \\
-15 & 14.68$\pm$0.03 &  8.63$\pm$0.11 & -10.63$\pm$0.12 & -2.58$\pm$0.07 & 0.87$\pm$0.11 & 0.49$\pm$0.06 \\
 -5 & 14.80$\pm$0.03 &  9.10$\pm$0.11 &  -9.38$\pm$0.19 & -2.16$\pm$0.11 & 1.06$\pm$0.17 & 0.56$\pm$0.10 \\
  5 & 15.22$\pm$0.02 & 11.07$\pm$0.08 &  -7.66$\pm$0.25 & -1.82$\pm$0.21 & 1.37$\pm$0.23 & 0.98$\pm$0.18 \\
 15 & 15.41$\pm$0.01 & 12.06$\pm$0.08 &  -6.33$\pm$0.40 & -1.77$\pm$0.40 & 1.44$\pm$0.33 & 1.58$\pm$0.36 \\
 25 & 15.98$\pm$0.01 & 15.73$\pm$0.05 &  -4.64$\pm$0.35 & -1.66$\pm$0.36 & 1.31$\pm$0.29 & 1.46$\pm$0.31 \\
\hline 
\end{tabular}
\end{center}
\label{GD1vals}
\end{table*}

\subsection{GD-1 proper motions}\label{proper_motions}
With distances to the stream in place, we now revisit the PM selection for GD-1 stars. We once again select the sample of probable GD-1 stars consistent with the reference isochrone within twice the photometric errors, but at the distance as given by Figure~\ref{GD1_distfit}. The PM cloud in $\mu_{\mathrm{\varphi_{1}}}$, $\mu_{\mathrm{\varphi_{2}}}$ is then fit using a Gaussian mixture model consisting of one Gaussian for the stream distribution and another for the Milky Way foreground distribution. Distributions are fit using the $\texttt{emcee}$ python MCMC package for each bin of $\varphi_{1}$~\citep{emcee}. 
\begin{figure}
\centering
\includegraphics[angle=0, width=0.495\textwidth]{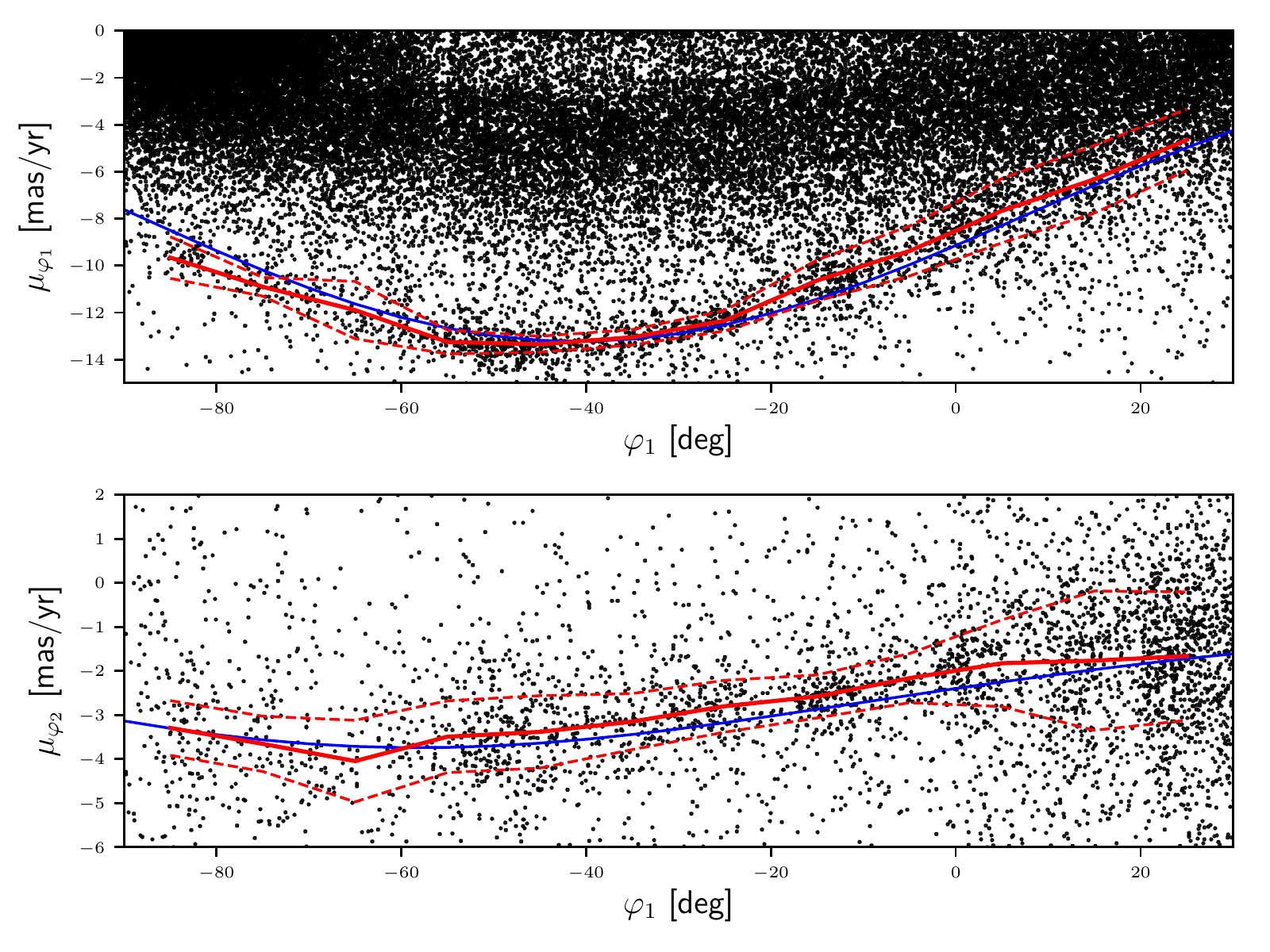}
\caption{Proper motion of GD-1 stars in stream-aligned coordinates $\varphi_{1},\varphi_{2}$ adopting the distances shown in Figure~\ref{GD1_distfit}. In the bottom panel, $\mu_{\mathrm{\varphi_{2}}}$ is shown only for stars with $\mu_{\mathrm{\varphi_{1}}}$ consistent with the best-fit within 2.5 mas/yr to better bring out the stream signal. The solid red line shows the best-fit PM solution as function of $\varphi_{1}$ for $\mu_{\mathrm{\varphi_{1}}}$ in the top panel and $\mu_{\mathrm{\varphi_{2}}}$ in the bottom, with dashed lines indicating the Gaussian standard deviation. The blue line shows the PM trends from a simulation aimed to reproduce the observed properties of GD-1 (Erkal et al., in prep). \label{GD1_PMfit}}
\end{figure}

Figure~\ref{GD1_PMfit} shows the extracted samples of GD-1 stars, with $\mu_{\mathrm{\varphi_{1}}}$ in the top panel and $\mu_{\mathrm{\varphi_{2}}}$ in the bottom. For reference, the PMs from a simulation aimed to reproduce the observed properties of GD-1 are shown as the blue line, showing good agreement (see Section~\ref{sec:orbits} for more details). The stream signal is readily visible in $\mu_{\mathrm{\varphi_{1}}}$, with a good separation from the MW PMs for -90$<\varphi_{1}<$0 deg. Beyond $\varphi_{1}$=0 deg, the stream PMs become similar to those of the MW, leading to increased contamination of the stream sample. The stream PMs in $\varphi_{2}$ are separated as much from those of the MW as for $\mu_{\mathrm{\varphi_{1}}}$. Therefore, in the bottom panel of Figure~\ref{GD1_PMfit}, we only show stars with $\mu_{\mathrm{\varphi_{1}}}$ consistent with our best-fit within 2.5 mas/yr to better bring out the stream signal. The GD-1 stars are visibly overdense in $\mu_{\mathrm{\varphi_{2}}}$ but not as distinct from the MW distribution as in $\varphi_{1}$ coordinates. Solid red lines in Figure~\ref{GD1_PMfit} shows the best-fit solution as function of $\varphi_{1}$, with dashed lines indicating the Gaussian standard deviation. The best-fit PM values of GD-1 are also shown in table~\ref{GD1vals}.

\section{Fitting the stream track of GD-1}\label{stream_track}
Next up, we will constrain the stream track of GD-1 on the sky, by combining simple PM selections (similar to what was done by \citealt{Price-Whelan18a}) with CMD matched filtering. In this way, we will extract the optimum information from the combined {\it Gaia} DR2 and Pan-STARRS dataset. Matched filtering relies on selecting an appropriate signal filter P$_{\mathrm{str}}$ for the population of interest (the stream), and comparing that to a suitable background population filter P$_{\mathrm{BG}}$ to find which pixels of a CMD provide the optimum signal-to-noise of the signal population \citep[see e.g.][]{Rockosi02}. In our case, we use the reference Padova isochrone with [Fe/H]=-1.9, age=12 Gyr as our stream population template, and generate a synthetic CMD by drawing stars from the isochrone with masses following a Kroupa IMF \citep{Kroupa01}. The colours and magnitudes of the synthetic stars are shifted to the appropriate distance for each $\varphi_{1}$ considered (using the function determined in Section~\ref{distance}) and convolved with photometric errors for that spatial bin to create a signal filter consistent with the observational conditions of the observed stars. To avoid a filter that is arbitrarily thin at the bright end (where photometric errors are small), a minimum filter magnitude width of 0.05 is adopted.
\begin{figure*}
\centering
\includegraphics[angle=0, width=0.95\textwidth]{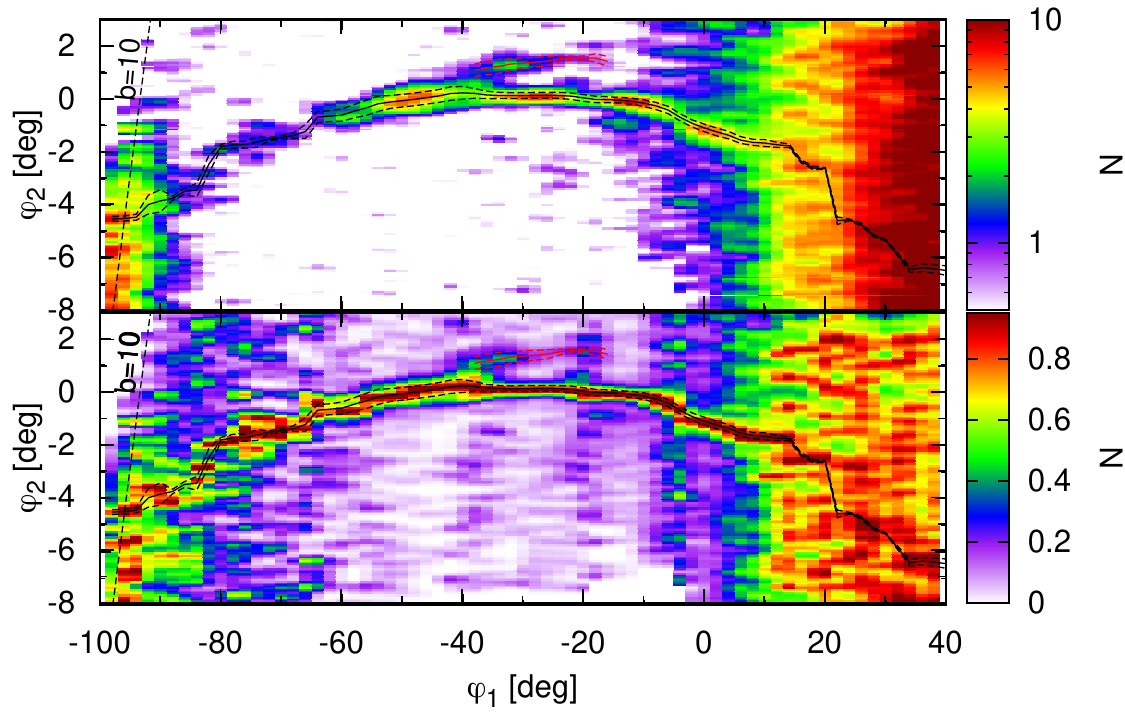}
\caption{Matched filter map in GD-1 centric coordinates after applying the distance and PM relations shown in Figures~\ref{GD1_distfit} and~\ref{GD1_PMfit}, using spatial bins of 2$\times$0.1 deg bins. The top panel shows a convolved matched filter density map, while the bottom panel shows a column normalised version of the convolved map to increase the stream track contrast. The maps have been convolved using a Gaussian kernel of 1x1 pixel to smooth over stochastic noise. The black lines show the fitted stream track (and widths), while the red lines show the fit to the spur track. \label{GD1_Gaia_trackfit}}
\end{figure*}

The background filter is constructed by selecting observed stars far enough away from the track locus defined by \citep{Price-Whelan18a} to be free of stream stars. We include only stars with $\varphi_{2}$ greater than 1.5 and less than 3 degrees away from the track locus on each side, and apply a PM cut to within 2 mas/yr of the stream motions following Section~\ref{proper_motions}. To ensure our background filter samples the same mix of MW populations as the signal filter, we select only stars with Galactic latitudes consistent with the mean latitude of the signal filter pixel, to within 5 degrees. The resulting background sample is converted into a Hess and further convolved using a 2x2 pixel kernel to smoothen the MW populations and avoid spurious hot pixels in the signal-to-noise filtering due to low number sampling. To exclude selecting other known streams that cross the same area of sky, we also include an extra background filter during the matched filter process, based on the stellar populations (and distance) of the Orphan and Sagittarius streams. This ensures that CMD bins coincident with these streams are rejected in the filtering, thereby avoiding contamination from other streams.

Both of the filters are then normalised and the fraction P$_{\mathrm{str}}$(g-r,r)/P$_{\mathrm{BG}}$(g-r,r) is used to determine the filter that provide the optimum stream detection signal-to-noise. In our case, we adopt a boolean matched filter procedure instead of a CMD pixel weighting procedure, since that will preserve the Poisson distribution for the data and allow for realistic error determination. We simply assign a weight of zero to values of P$_{\mathrm{str}}$(g-r,r)/P$_{\mathrm{BG}}$(g-r,r) lower than the adopted threshold and a weight of 1 for value higher than the threshold. The optimum threshold is chosen for each spatial pixel by looping through the possible values and finding the one which maximises the total signal-to-noise in the CMD. 

The resulting matched filter map of the GD-1 stream is shown in Figure~\ref{GD1_Gaia_trackfit}. The map is convolved using a Gaussian kernel of 1x1 pixels to smooth over stochastic noise and bring out low contrast stream features in more detail. The top panel shows the convolved map, with pixel colours indicating the number of stream stars, while the bottom panel shows a column normalised version of the map aimed at highlighting the stream track locus. The stream is readily visible, as in \citet{Price-Whelan18a}, and spans a significant range in $\varphi_{1}$, extending well beyond the footprint shown in SDSS data \citep{Grillmair061}. Figure~\ref{GD1_Gaia_trackfit} shows that the stream can be reliably traced between -85$<\varphi_{1}<$+15 deg before being swamped by contamination. This is due in part to the increase in MW disk density on the left side of the coverage as the stream approaches the MW mid plane, and the . However, another reason for losing the stream signal is due to the increasing distance on either end of the stream coverage (see Figure~\ref{GD1_distfit}). For distances greater than 10 kpc, there are not enough stars above the {\it Gaia} brightness limit to trace the stream confidently. Therefore, it might be possible to trace the GD-1 stream further by obtaining deeper data. While deeper PM data is ideal, suitably deep photometry would be sufficient to allow for good separation between the stream signal and background MW populations, as shown in \citet{deBoer2018a}.

To extract the GD-1 stream track, we fit a Gaussian density model and background to each column of $\varphi_{1}$ data. The recovered stream track locus is shown as the solid black line in Figure~\ref{GD1_Gaia_trackfit}, with the stream width indicated by dashed lines. The values for the track centre and width are also listed in Table~\ref{GD1_track_dens}. The stream track shows noticeable wiggles and changes in the width are also readily apparent in Figure~\ref{GD1_Gaia_trackfit}, most notably the increasing width toward lower $\varphi_{1}$. The recovered track is mostly consistent with the track determined from deep photometric data in \citet{deBoer2018a}, given the uncertainties. The difference in the track location is most pronounced in the region covering $-45<\varphi_{1}<-36$ deg, and likely influenced by the spur feature connecting on to the main stream track.

Given the relevance of the spur feature for the formation of the stream, we also constrain the track of the spur itself, by repeating the fitting in the spatial region showing the spur ($-40<\varphi_{1}<-15$ deg) using a double Gaussian fit. The recovered parameters for the spur track are shown in Table~\ref{GD1_track_dens_spur}. We stress that the fitting of two simultaneous Gaussian profile does not lead to changes in the nominal stream track, which is well separated from the spur in the regions studied. The spur track is shown in Figure~\ref{GD1_Gaia_trackfit} as a red line along with dashed lines indicating the recovered width. 

\begin{figure}
\centering
\includegraphics[angle=0, width=0.5\textwidth]{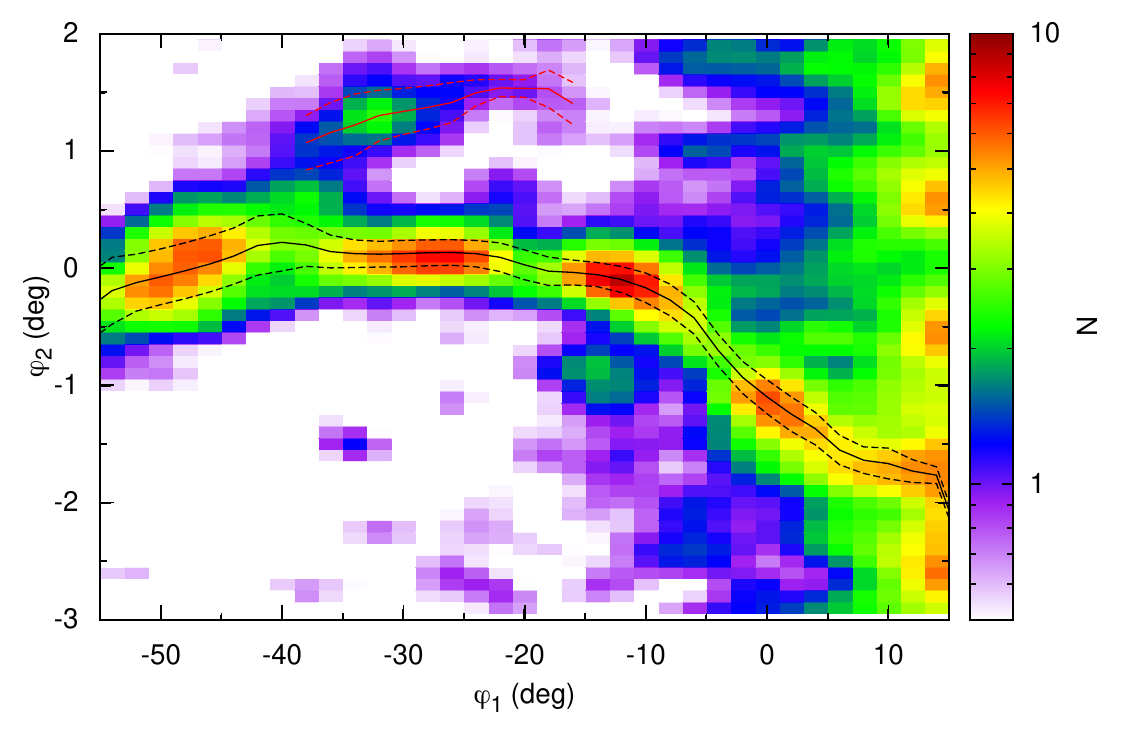}
\caption{A zoom of the matched filter map shown in Figure~\ref{GD1_Gaia_trackfit}, highlighting the region of the spur and blob discovered in \citet{Price-Whelan18a}. The black lines show the fitted stream track (and widths), while the red lines show the fit to the spur track. \label{GD1_Gaia_trackzoom}}
\end{figure}

To discuss the stream morphology in more detail, Figure~\ref{GD1_Gaia_trackzoom} shows a zoom of the convolved density map highlighting some of the more interesting regions of the stream. Once again, the stream clearly shows wiggles, similar to those found in \citet{deBoer2018a}. Three clear underdense regions can be identified, at $\varphi_{1}\approx$-36,-20 and -3 deg respectively. The stellar spur coming off the stream is clearly visible and seen to extend slightly further and farther out than found in \citet{Price-Whelan18a}. The feature also appears to arc back toward the stream track at $\varphi_{1}\approx$-20 deg, although the stellar sample there is too small to be significant, and needs to be followed up with deeper data. Interestingly, we find that the spur extends over most of the adjacent gap at $\varphi_1 \sim -36$ deg instead of connecting on to the edge of the gap as expected \cite[e.g.][]{Erkal15b}. This is also consistent with the deep photometric study performed in \citet{deBoer2018a} as visible in their Figure 8. We will discuss the interpretation of this in more detail in Section~\ref{sec:spur_morphology}. 

The blob feature from \citet{Price-Whelan18a} is also visible in Figure~\ref{GD1_Gaia_trackzoom} at $[\varphi_{1},\varphi_{2}] \approx [-14.7,1.0]$ deg ([RA,Dec] $\approx$ [176.3,52.7]), composed of $\approx$70 stars. The feature covers roughly 8 deg along the stream and cannot be connected to the stream with the present data. Finally, we find an underdense feature at $\varphi_{1}\approx-3$ deg, which was neither seen in data from \citet{deBoer2018a} nor \citet{Price-Whelan18a} but was detected in \citet{malhan_2019}. This feature is especially intriguing given the distinct sinusoidal wiggles in the stream track on each side of the under density. We argue that this feature is likely a perturbation to the stream in  Section~\ref{sec:wiggle}.

\begin{table}
\caption[]{The track centre $\varphi_{2}$, width $\sigma_{\mathrm{\varphi_{2}}}$, surface brightness I, linear density $\mathrm{\lambda}$ and mass sampling fraction ($_{\mathrm{samp}}$ of the GD-1 stream as a function of $\varphi_{1}$.}
\begin{center}
\begin{tabular}{cccrcc}
\hline\hline
$\varphi_{1}$ &  $\varphi_{2,track}$ & $\sigma_{\mathrm{\varphi_{2}}}$ & \multicolumn{1}{c}{\centering I}  & $\mathrm{\lambda}$ & f$_{\mathrm{samp}}$ \\
{[deg]} & [deg] & [deg] & \multicolumn{1}{c}{\centering[deg$^{-2}$]} & [deg$^{-1}$] &  \\
\hline
-88 & -3.74$\pm$0.16 & 0.08$\pm$0.02 &  7.28$\pm$1.51 & 0.77$\pm$0.20 & 0.83 \\
-86 & -3.47$\pm$0.19 & 0.11$\pm$0.03 &  5.73$\pm$1.14 & 0.78$\pm$0.25 & 0.86 \\
-84 & -3.43$\pm$0.68 & 0.26$\pm$0.12 &  2.76$\pm$0.86 & 0.88$\pm$0.41 & 0.88 \\
-82 & -2.46$\pm$0.71 & 0.33$\pm$0.17 &  2.22$\pm$0.66 & 0.93$\pm$0.43 & 0.89 \\
-80 & -1.80$\pm$0.23 & 0.09$\pm$0.02 &  8.09$\pm$1.73 & 0.90$\pm$0.25 & 0.91 \\
-78 & -1.73$\pm$0.10 & 0.13$\pm$0.03 &  6.92$\pm$1.30 & 1.12$\pm$0.29 & 0.92 \\
-76 & -1.72$\pm$0.08 & 0.13$\pm$0.04 &  5.80$\pm$1.25 & 0.94$\pm$0.28 & 0.94 \\
-74 & -1.62$\pm$0.08 & 0.20$\pm$0.09 &  3.74$\pm$0.87 & 0.94$\pm$0.37 & 0.95 \\
-72 & -1.51$\pm$0.07 & 0.06$\pm$0.05 &  3.68$\pm$1.91 & 0.28$\pm$0.22 & 0.96 \\
-70 & -1.45$\pm$0.06 & 0.07$\pm$0.03 &  5.58$\pm$1.90 & 0.48$\pm$0.23 & 0.96 \\
-68 & -1.38$\pm$0.07 & 0.10$\pm$0.04 &  4.91$\pm$1.46 & 0.65$\pm$0.27 & 0.98 \\
-66 & -1.22$\pm$0.14 & 0.26$\pm$0.08 &  4.75$\pm$0.74 & 1.56$\pm$0.41 & 0.98 \\
-64 & -0.69$\pm$0.07 & 0.22$\pm$0.07 &  4.77$\pm$0.91 & 1.29$\pm$0.39 & 0.99 \\
-62 & -0.65$\pm$0.05 & 0.20$\pm$0.05 &  5.79$\pm$1.07 & 1.48$\pm$0.40 & 0.99 \\
-60 & -0.61$\pm$0.05 & 0.23$\pm$0.05 &  7.92$\pm$1.10 & 2.31$\pm$0.51 & 1.00 \\
-58 & -0.47$\pm$0.05 & 0.25$\pm$0.04 & 11.46$\pm$1.22 & 3.54$\pm$0.61 & 1.00 \\
-56 & -0.34$\pm$0.05 & 0.29$\pm$0.04 & 14.83$\pm$1.16 & 5.33$\pm$0.70 & 1.00 \\
-54 & -0.19$\pm$0.03 & 0.29$\pm$0.03 & 18.55$\pm$1.17 & 6.64$\pm$0.69 & 1.01 \\
-52 & -0.12$\pm$0.02 & 0.24$\pm$0.02 & 23.71$\pm$1.51 & 7.22$\pm$0.64 & 1.01 \\
-50 & -0.07$\pm$0.02 & 0.24$\pm$0.01 & 27.83$\pm$1.66 & 8.28$\pm$0.69 & 1.01 \\
-48 & -0.02$\pm$0.02 & 0.24$\pm$0.02 & 29.30$\pm$1.70 & 8.77$\pm$0.73 & 1.00 \\
-46 &  0.04$\pm$0.03 & 0.24$\pm$0.02 & 27.23$\pm$1.64 & 8.15$\pm$0.71 & 1.01 \\
-44 &  0.10$\pm$0.03 & 0.24$\pm$0.02 & 22.33$\pm$1.50 & 6.73$\pm$0.66 & 1.01 \\
-42 &  0.20$\pm$0.03 & 0.25$\pm$0.03 & 16.99$\pm$1.32 & 5.38$\pm$0.65 & 1.01 \\
-40 &  0.22$\pm$0.04 & 0.24$\pm$0.03 & 13.78$\pm$1.29 & 4.20$\pm$0.61 & 1.01 \\
-38 &  0.20$\pm$0.04 & 0.18$\pm$0.03 & 13.59$\pm$1.54 & 3.12$\pm$0.50 & 1.00 \\
-36 &  0.14$\pm$0.03 & 0.14$\pm$0.02 & 16.41$\pm$1.93 & 2.87$\pm$0.46 & 1.00 \\
-34 &  0.13$\pm$0.02 & 0.12$\pm$0.01 & 20.58$\pm$2.39 & 3.02$\pm$0.47 & 0.99 \\
-32 &  0.12$\pm$0.02 & 0.11$\pm$0.01 & 25.88$\pm$2.74 & 3.53$\pm$0.49 & 0.98 \\
-30 &  0.13$\pm$0.02 & 0.11$\pm$0.01 & 30.54$\pm$2.85 & 4.30$\pm$0.52 & 0.98 \\
-28 &  0.13$\pm$0.02 & 0.11$\pm$0.01 & 35.13$\pm$2.99 & 4.87$\pm$0.53 & 0.97 \\
-26 &  0.14$\pm$0.02 & 0.11$\pm$0.01 & 36.05$\pm$3.03 & 4.90$\pm$0.52 & 0.96 \\
-24 &  0.13$\pm$0.02 & 0.11$\pm$0.01 & 29.82$\pm$2.78 & 4.20$\pm$0.50 & 0.95 \\
-22 &  0.10$\pm$0.03 & 0.12$\pm$0.01 & 20.71$\pm$2.31 & 3.17$\pm$0.47 & 0.94 \\
-20 &  0.03$\pm$0.03 & 0.13$\pm$0.02 & 15.94$\pm$2.01 & 2.54$\pm$0.44 & 0.93 \\
-18 & -0.02$\pm$0.03 & 0.12$\pm$0.01 & 18.45$\pm$2.24 & 2.79$\pm$0.46 & 0.92 \\
-16 & -0.03$\pm$0.02 & 0.11$\pm$0.01 & 28.42$\pm$2.91 & 3.80$\pm$0.52 & 0.90 \\
-14 & -0.05$\pm$0.02 & 0.10$\pm$0.01 & 38.30$\pm$3.42 & 4.92$\pm$0.58 & 0.89 \\
-12 & -0.09$\pm$0.02 & 0.11$\pm$0.01 & 40.02$\pm$3.36 & 5.59$\pm$0.62 & 0.87 \\
-10 & -0.16$\pm$0.02 & 0.13$\pm$0.01 & 35.80$\pm$2.97 & 5.63$\pm$0.63 & 0.85 \\
 -8 & -0.27$\pm$0.03 & 0.13$\pm$0.01 & 29.76$\pm$2.60 & 5.01$\pm$0.59 & 0.83 \\
 -6 & -0.42$\pm$0.05 & 0.14$\pm$0.01 & 23.57$\pm$2.38 & 4.06$\pm$0.56 & 0.81 \\
 -4 & -0.70$\pm$0.06 & 0.14$\pm$0.02 & 20.18$\pm$2.35 & 3.42$\pm$0.56 & 0.79 \\
 -2 & -0.93$\pm$0.05 & 0.14$\pm$0.02 & 21.50$\pm$2.47 & 3.68$\pm$0.60 & 0.77 \\
  0 & -1.09$\pm$0.04 & 0.15$\pm$0.02 & 24.40$\pm$2.54 & 4.45$\pm$0.67 & 0.75 \\
  2 & -1.24$\pm$0.04 & 0.15$\pm$0.02 & 24.59$\pm$2.59 & 4.53$\pm$0.69 & 0.73 \\
  4 & -1.37$\pm$0.05 & 0.14$\pm$0.02 & 21.34$\pm$2.65 & 3.71$\pm$0.67 & 0.71 \\
  6 & -1.55$\pm$0.05 & 0.13$\pm$0.03 & 16.64$\pm$2.70 & 2.64$\pm$0.63 & 0.69 \\
  8 & -1.64$\pm$0.04 & 0.11$\pm$0.04 & 12.57$\pm$2.78 & 1.79$\pm$0.60 & 0.67 \\
 10 & -1.66$\pm$0.05 & 0.13$\pm$0.06 & 13.27$\pm$3.40 & 2.19$\pm$0.97 & 0.63 \\
 12 & -1.73$\pm$0.05 & 0.10$\pm$0.05 &  9.48$\pm$3.22 & 1.16$\pm$0.62 & 0.60 \\
 14 & -1.76$\pm$0.07 & 0.07$\pm$0.04 & 12.08$\pm$4.12 & 1.07$\pm$0.56 & 0.59 \\
 16 & -2.35$\pm$0.18 & 0.06$\pm$0.03 & 12.40$\pm$4.48 & 0.94$\pm$0.51 & 0.55 \\
 18 & -2.59$\pm$0.07 & 0.06$\pm$0.04 & 10.51$\pm$4.25 & 0.82$\pm$0.51 & 0.53 \\
 20 & -2.62$\pm$0.06 & 0.04$\pm$0.62 &  8.06$\pm$5.22 & 0.35$\pm$0.47 & 0.51 \\
 22 & -4.58$\pm$0.07 & 0.14$\pm$0.31 &  5.03$\pm$7.45 & 0.89$\pm$0.71 & 0.48 \\
 24 & -4.55$\pm$0.09 & 0.03$\pm$0.04 &  8.74$\pm$6.67 & 0.33$\pm$0.40 & 0.46 \\
 26 & -4.73$\pm$0.27 & 0.06$\pm$0.06 & 11.70$\pm$6.85 & 0.89$\pm$0.82 & 0.43 \\
 28 & -5.14$\pm$0.20 & 0.05$\pm$0.04 & 13.67$\pm$6.84 & 0.79$\pm$0.65 & 0.41 \\
 30 & -5.31$\pm$0.21 & 0.03$\pm$0.03 & 11.38$\pm$7.83 & 0.41$\pm$0.42 & 0.39 \\
\hline 
\end{tabular}
\end{center}
\label{GD1_track_dens}
\end{table}

\section{Stream density and total mass}\label{stream_density}
\begin{figure*}
\centering
\includegraphics[angle=0, width=0.995\textwidth]{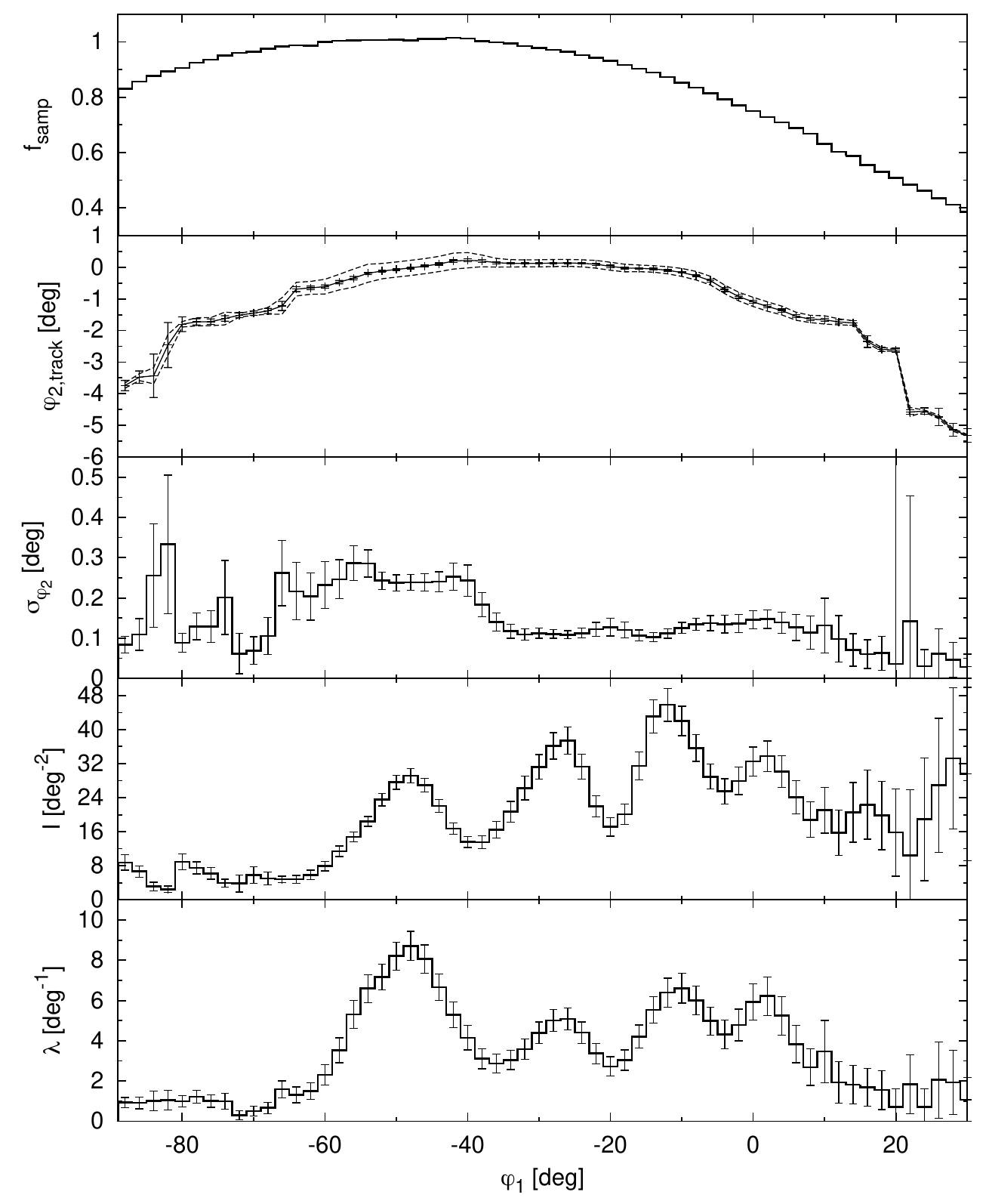}
\caption{Overview of the best-fit parameters of the GD-1 stream, showing the mass sampling fraction (f$_{\mathrm{samp}}$), track position $\varphi_{2,track}$ (with stream width shown as dashed lines), stream width ($\sigma_{\mathrm{\varphi_{2}}}$), surface brightness I$(\mathrm{\varphi_{1}})$ and linear density $\mathrm{\lambda}(\mathrm{\varphi_{1}})$. Parameters have been determined by fitting a Gaussian plus 1st order polynomial background to the data shown in Figure~\ref{GD1_Gaia_trackfit}. The surface brightness and linear density shown here have been corrected for the effects of mass sampling given in the top panel. \label{GD1_stream_info}}
\end{figure*}

Following the stream track determination, we will now constrain the surface brightness and density of the GD-1 stream.  This will reveal the location and depth of the under densities shown in Figure~\ref{GD1_Gaia_trackfit} as well as the overall density evolution of the stream. To determine the stream densities, we construct another matched filter map of GD-1 using a finer binning in a transformed coordinates $\varphi_{1}$,$\hat{\varphi_{2}}$, in which $\hat{\varphi_{2}}=\varphi_{2}-\varphi_{2,0}$ and $\varphi_{2,0}$ being the stream track location for each star interpolated using the values given in Table~\ref{GD1_track_dens}. The resulting number densities are then fit using a Gaussian profile plus a first order polynomial for the residual background contamination. The fits allow us to determine the surface brightness I$(\mathrm{\varphi_{1}})$ of the GD-1 stream as a function of angle along the stream as well as the linear density $\mathrm{\lambda}(\mathrm{\varphi_{1}})$, calculated as $\mathrm{\lambda}(\mathrm{\varphi_{1}})=\sqrt{2\pi}$I$(\mathrm{\varphi_{1}})$$\sigma_{\mathrm{\varphi_{2}}}$. The resulting values are shown in Figure~\ref{GD1_stream_info}, along with the stream track location $\varphi_{2,{\rm track}}$ and stream width $\sigma_{\mathrm{\varphi_{2}}}$. We note that the recovered distance, track location and proper motions are in very good agreement with the dynamical model of GD-1 by \citet{Bovy16}, even outside the range of $\varphi_{1}$ probed in the SDSS data modelled there, making it the model in the literature that best reproduces the behaviour of the stream track. The surface brightness and linear density of the spur are also computed using the same procedure, given the track determined in Section~\ref{stream_track}, and shown in Table~\ref{GD1_track_dens_spur}.

\begin{table}
\caption[]{The track centre $\varphi_{2}$, width $\sigma_{\mathrm{\varphi_{2}}}$, surface brightness I, linear density $\mathrm{\lambda}$ of the stream spur as a function of $\varphi_{1}$.}
\begin{center}
\begin{tabular}{cccrc}
\hline\hline
$\varphi_{1}$ &  $\varphi_{2,track}$ & $\sigma_{\mathrm{\varphi_{2}}}$ & \multicolumn{1}{c}{\centering I}  & $\mathrm{\lambda}$  \\
{[deg]} & [deg] & [deg] & \multicolumn{1}{c}{\centering[deg$^{-2}$]} & [deg$^{-1}$]  \\
\hline
-38 & 1.07$\pm$0.20 & 0.23$\pm$0.14 & 5.24$\pm$1.41 & 1.53$\pm$0.79 \\
-36 & 1.16$\pm$0.07 & 0.26$\pm$0.10 & 6.34$\pm$1.19 & 2.04$\pm$0.69 \\
-34 & 1.22$\pm$0.07 & 0.26$\pm$0.10 & 7.22$\pm$1.23 & 2.40$\pm$0.78 \\
-32 & 1.30$\pm$0.05 & 0.22$\pm$0.07 & 8.03$\pm$1.44 & 2.18$\pm$0.61 \\
-30 & 1.34$\pm$0.05 & 0.20$\pm$0.06 & 7.66$\pm$1.46 & 1.89$\pm$0.52 \\
-28 & 1.37$\pm$0.06 & 0.18$\pm$0.06 & 6.09$\pm$1.37 & 1.41$\pm$0.46 \\
-26 & 1.41$\pm$0.08 & 0.17$\pm$0.08 & 4.24$\pm$1.31 & 0.91$\pm$0.42 \\
-24 & 1.50$\pm$0.08 & 0.11$\pm$0.08 & 3.22$\pm$1.62 & 0.45$\pm$0.33 \\
-22 & 1.54$\pm$0.09 & 0.07$\pm$0.09 & 2.44$\pm$2.02 & 0.22$\pm$0.26 \\
-20 & 1.54$\pm$0.12 & 0.07$\pm$0.12 & 1.80$\pm$1.87 & 0.17$\pm$0.26 \\
-18 & 1.53$\pm$0.12 & 0.16$\pm$0.21 & 1.83$\pm$1.64 & 0.37$\pm$0.48 \\
-16 & 1.41$\pm$0.18 & 0.18$\pm$0.35 & 3.46$\pm$2.38 & 1.22$\pm$1.35 \\
\hline 
\end{tabular}
\end{center}
\label{GD1_track_dens_spur}
\end{table}

Given the variation of stream distance as a function of $\varphi_{1}$ as determined in Section~\ref{distance}, the range of masses sampled is not equal across the covered footprint. To correct for this potential biasing factor in the densities, we compute the total mass sampled above the {\it Gaia} DR2 brightness limit, in each bin of $\varphi_{1}$. The masses of the synthetic GD-1 stars used to construct the signal filter (see Section~\ref{stream_track}) are combined with the optimised matched filter to find the total stellar mass sampled at each $\varphi_{1}$ given the distance and photometric errors (controlling the filter width). Following this, we derive the relative mass sampling fraction f$_{\mathrm{samp}}$ in each $\varphi_{1}$ bin relative to the bin at $\varphi_{1}$=-40 deg. The resulting mass sampling fractions are given in table~\ref{GD1_track_dens} and shown in Figure~\ref{GD1_stream_info}. Over most of the extent of the GD-1 stream probed here, f$_{\mathrm{samp}}$ is consistent to within 20\%, showing only a limited effect on the density determination. However, for $\varphi_{1}>$-10 deg the {\it Gaia} DR2 data probes less and less of the stream as the distance increases (see Figure~\ref{GD1_distfit}), culminating in a 40\% sampling at $\varphi_{1}$=30 deg. In Figure~\ref{GD1_stream_info}, we have corrected the surface brightness I$_{\mathrm{\varphi_{1}}}$ and linear density $\mathrm{\lambda}_{\mathrm{\varphi_{1}}}$ for the relative effects of mass sampling shown in the top panel.

The surface brightness and linear density displayed in Figure~\ref{GD1_stream_info} make it clear that the GD-1 stream has experienced significant density fluctuations across the extent studied here. There are four clear density peaks separated by three under densities of varying depth. Besides the peaks, the more negative $\varphi_{1}$ side of GD-1 shows a decreasing density accompanied by increasing stream width until the stream signal is lost. On the positive $\varphi_{1}$ side the linear density also decreases, which is partially driven by a narrowing of the stream instead of a lower surface brightness. Given that is the most distant part of the stream sampled, the uncertainties on the density are highest here.

The morphology of GD-1 complicates determining which of the under densities is most significant, since there is no clear stream continuum level and no progenitor. Nonetheless, we can conclude that the highest density part of the GD-1 stream is centered around $\varphi_{1}$=-48 deg, close to the location of the spur feature and a possible place where it attaches to the main stream track.  The three other peaks at $\varphi_{1}$=-27,-10 and +2 deg are lower by nearly a factor two. The three under densities are located at $\varphi_{1}$=-36,-20 and -3 deg, with the one at most negative $\varphi_{1}$ showing a different location ($\varphi_{1}$=-38 deg) in surface brightness compared to linear density. This might be related to the spur feature, which is expected to influence the stream width in this region. The least pronounced under density at $\varphi_{1}$=-3 deg is flanked by similar density peaks, shown to display a sinusoidal wiggle in the track in Figure~\ref{GD1_Gaia_trackfit}. While this wiggle looks similar to characteristic ``S"-shape of stars coming off the stream's progenitor \citep[as in Pal 5,][]{pal5_disc}, this wiggle has the wrong orientation given the orbit of GD-1 \citep[e.g. Fig. 12 of][]{deBoer2018a}. Thus, this feature must instead come from a perturbation to the stream. Since GD-1 is on a retrograde orbit, the effect of baryonic substructure in the disk should be minimal \citep{Amorisco16} and thus this is a promising candidate for a subhalo interaction. Deeper data is needed to unambiguously determine the stream density here and to confirm the wiggle in the stream. We discuss this feature in more detail in Section~\ref{sec:wiggle}.  

Besides determining the mass sampling fraction in each $\varphi_{1}$ bin, we can also use the synthetic CMDs to determine the total mass and number of stars in the sampled region of the stream. Similar to the method described above, we can determine what fraction of the total mass is sampled within the optimised matched filter. Therefore, we can convert the observed densities to the total initial number and mass of the stream.

We start of with a fully sampled population of stars within a mass range of 0.1-120 M$_{\odot}$, under the assumption of a Kroupa IMF \citep{Kroupa01}. We then use an isochrone with the appropriate age and metallicity to generate the number/masses of stars that are still alive at the present day. These synthetic stars are then convolved with observational effects and the optimised matched filter to get the filtered, current day density of stars that this population corresponds to. This number can be directly compared to the observed densities to scale the initial mass of the population up or down, and determine the initial mass and initial number of stars of the stream segment. The obtained number is computed under the assumption of a mass range of 0.1-120 M$_{\odot}$ and Kroupa IMF, and  will be different if there are changes in some of the more uncertain evolutionary factors such as mass loss or unusual stellar evolution processes. Given these restrictions, we determine that the visible stretch of GD-1 corresponds to a total of $2.6\pm0.09\times10^{4}$ initial stars, with a total initial stellar mass of $1.80\pm0.13\times10^{4}~{\rm M}_{\odot}$. 

We also determine the total initial stellar mass contained within the spur feature within the spatial region over which we can determine its track (see Section~\ref{stream_track}). Given these considerations, we find that the spur should contain a total of 1.17$\pm$0.15$\times$10$^{3}$ stars, with a total stellar mass of 0.81$\pm$0.22$\times$10$^{3}$ M$_{\odot}$.
\begin{figure*}
\centering
\includegraphics[angle=0, width=0.995\textwidth]{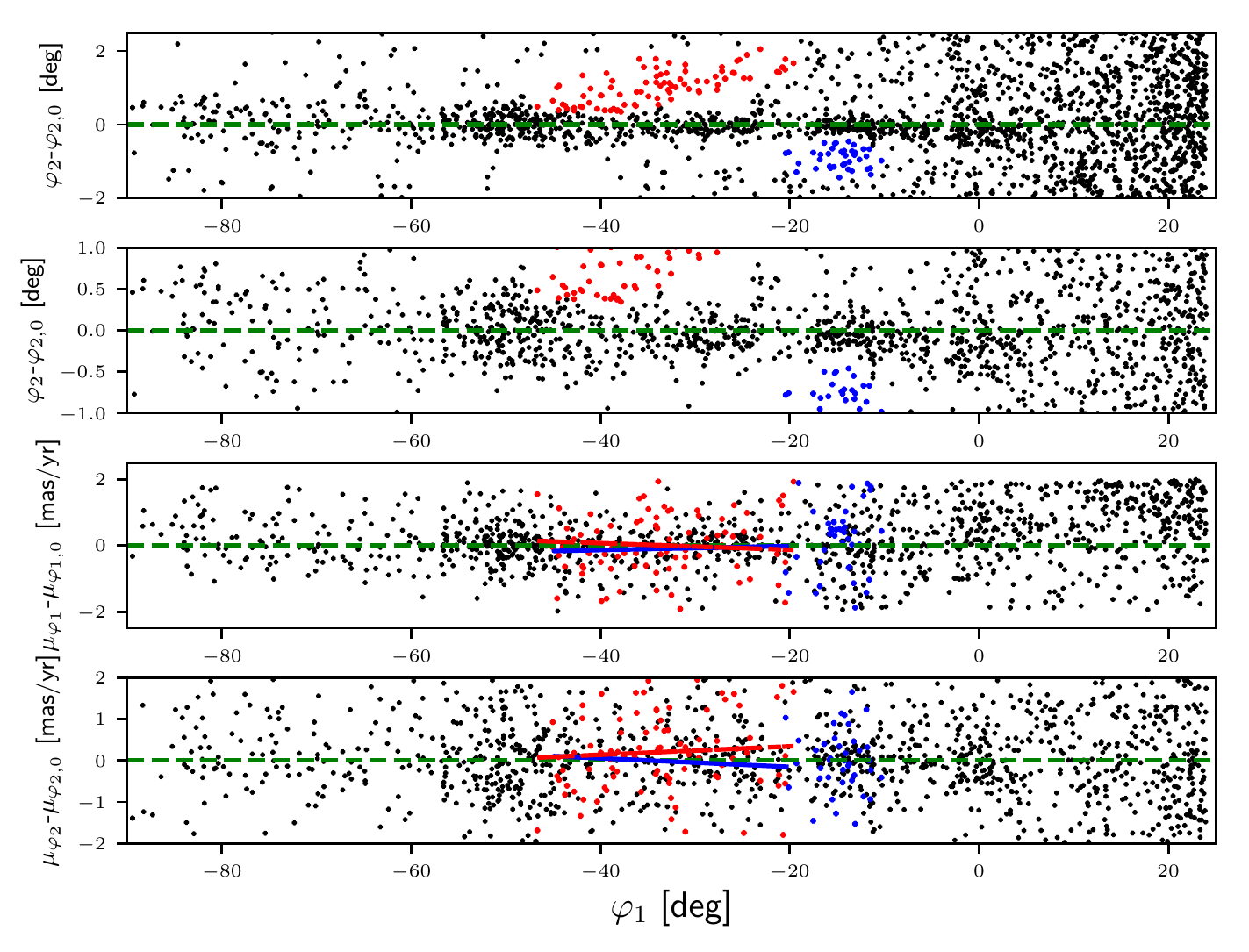}
\caption{Residuals of the extracted member stars of the GD-1 stream, in position (top panel) with zoom-in (second panel) and PM space (bottom two panels). Black points indicate the overall stream sample, while red (blue) points show the spatially selected spur (blob) stars, given the rough positions from \citet{Price-Whelan18a} The lines show straight line fits to the proper motions of the spur and underlying  stream stars. \label{GD1_resids}}
\end{figure*}

Compared to other stellar streams, such as those found in the DES survey, GD-1 has an intermediate total stellar mass \citep{Shipp18}. However, the GD-1 stream is much longer than the streams found in the DES survey, while also being narrower across that length. Therefore, GD-1 is a very low density stream spanning a wide range across the sky, making it a very interesting object to try to reproduce in simulations. In particular, it would be extremely valuable to determine the formation timescale of a stream like GD-1, and the survival chances of low mass globular clusters in the MW across that timescale.

We can compare these numbers to the lifetimes of dissolving GCs from \citet{Baumgardt03}, who ran numerical simulations of dissolving GCs in realistic tidal fields. Using their equations 7 and 8, with a GD-1 apocentre distance of 28~kpc \citep{Willett09}, we find that a GC with an initial number of 1.3$\times$10$^{4}$ stars has a dissolution time of roughly 12 Gyr, being the upper limit of GD-1 dissolution. This corresponds to an initial mass of roughly 0.88$\times$10$^{4}$ M$_{\odot}$ and about half that at the present day after mass loss driven by stellar evolution (i.e. $\sim4.4\times10^3~{\rm M}_\odot$). The difference in mass is roughly a factor 4, with the stream having too much mass at the present day to be produced by the full dissolution of a GC within the appropriate time interval. This indicates that  either the progenitor of the GD-1 has not (fully) dissolved, or there is some missing physics in the \citet{Baumgardt03} simulations.  This discrepancy would be exacerbated if we consider that we are not probing the full length of the stream within the {\it Gaia} DR2 data. To produce higher dissolved stream masses, cases that boost the stellar escape rate need to be considered, such as starting with low density progenitors or retaining black holes within the parent GC \citep[][Erkal, Gieles, \& de Boer in prep]{Giersz2019}.

\section{Stream residuals}\label{residuals}
Besides the stream density and widths, we also look at the track and PM residuals along the stream, to search for hints of small scale wiggles and deviations. To facilitate this, we first extract the sample of stream stars from the data, given the parameters determined above. For each $\varphi_{1}$ bin, we extract all the stars contained within the maximum signal-to-noise matched filter and satisfying cuts in spatial position and PM around the best-fit tracks shown in Figures \ref{GD1_PMfit} and \ref{GD1_Gaia_trackfit}. Figure~\ref{GD1_resids} shows the residuals in both spatial coordinates $\hat{\varphi_{2}}=\varphi_{2}$-$\varphi_{2,0}$ and PMs $\mu_{\mathrm{\varphi_{1}}}-\mu_{\mathrm{\varphi_{1,0}}}$,$\mu_{\mathrm{\varphi_{2}}}-\mu_{\mathrm{\varphi_{2,0}}}$ for the extracted sample of GD-1 stars, as a function of $\varphi_{1}$. The second panel shows a zoom-in of the stream track to bring out small offsets in $\varphi_{2}$ around the stream track.

The stream is readily apparent as an over-density around the fitted tracks, although background contamination is also still visible. We have not attempted to quantify the stream membership probability here, since part of our goal is to investigate the behaviour of the features described in \citet{Price-Whelan18a}, which would be excluded on the basis of spatial membership probability. Figure~\ref{GD1_resids} makes it clear our fitted tracks well reproduce the large scale behaviour of GD-1 as function of $\varphi_{1}$. However, there are also signs of small scale wiggles in the stream track of order 0.2 deg, as already seen in \citet{deBoer2018a}. In particular, there is a small scale wiggle corresponding to the sinusoidal track shape shown in Figure~\ref{GD1_Gaia_trackzoom}, which does not appear to correspond to any clear changes in the PMs in the bottom two panels. 

Besides the main GD-1 stream stars, we have also highlighted the stars belonging to the spur and blob from \citet{Price-Whelan18a} in Figure~\ref{GD1_resids}. The selections of both features are based solely on the spatial positions, without considering selection in PM apart from the rough cuts around the track. The extracted stars are consistent with the stream PMs, showing unequivocally that these features are not due to background contamination. The width in PM space for both features appears slightly larger than for the underlying stream stars at similar $\varphi_{1}$. To study this in more detail, we determine the mean PM residuals and widths of both samples from the highlighted samples in Figure~\ref{GD1_resids}.

For the spur stars, the mean PM residuals are ($\mu_{\mathrm{\varphi_{1}}},\mu_{\mathrm{\varphi_{2}}})=(0.038,0.177)$ mas/yr along with standard deviation widths of ($\sigma_{\mu_{\mathrm{\varphi_{1}}}},\sigma_{\mu_{\mathrm{\varphi_{2}}}})=(0.882,0.927)$ mas/yr. This can be compared to the residuals for stars within the main stream (with $\vert\hat{\varphi_{2}}\vert<0.75$ deg), covering the same range of $\varphi_{1}$, which are ($\mu_{\mathrm{\varphi_{1}}},\mu_{\mathrm{\varphi_{2}}})=(-0.021,-0.004)$~mas/yr and widths ($\sigma_{\mu_{\mathrm{\varphi_{1}}}},\sigma_{\mu_{\mathrm{\varphi_{2}}}})=(0.620,0.802)$ mas/yr. Therefore, the spur stars are slightly offset from the stream and have a larger width in both $\mu_{\mathrm{\varphi_{1}}}$ and $\mu_{\mathrm{\varphi_{2}}}$. Notable, the proper motions of the spur stars point away from the stream in $\mu_{\mathrm{\varphi_{2}}}$. However, the sample sizes are small and the errors are large enough (the error on the mean is (0.096, 0.101) mas/yr for spur stars and (0.037,0.048) mas/yr for stream stars) that the differences are not significant enough to conclude unambiguously that the two samples show different PMs.

To investigate the proper motions of the spur in more detail, we divide the sample using a cut at $\varphi_{1}=-30$ degrees to select the spur closer to and further away from the possible connection point. For the spur below $\varphi_{1}=-30$ degrees, the mean PM residuals are ($\mu_{\mathrm{\varphi_{1}}},\mu_{\mathrm{\varphi_{2}}})=(-0.064,0.225)$ mas/yr along with standard deviation widths of ($\sigma_{\mu_{\mathrm{\varphi_{1}}}}$,$\sigma_{\mu_{\mathrm{\varphi_{2}}}})=(0.895,0.842)$ mas/yr. The error on the mean is (0.138,0.130) mas/yr. In comparison the stream stars covering the same spatial range show ($\mu_{\mathrm{\varphi_{1}}},\mu_{\mathrm{\varphi_{2}}})=(-0.090,0.022)$ mas/yr and widths ($\sigma_{\mu_{\mathrm{\varphi_{1}}}},\sigma_{\mu_{\mathrm{\varphi_{2}}}})=(0.649,0.774)$ mas/yr and an error on the mean of (0.069,0.082) mas/yr. Clearly, the proper motions are more like the stream closer to the connection point, and agree within roughly one sigma. However, for the spur above $\varphi_{1}=-30$ degrees things look slightly different.  The mean PM residuals of spur stars are ($\mu_{\mathrm{\varphi_{1}}},\mu_{\mathrm{\varphi_{2}}})=(0.162,0.335)$ mas/yr along with standard deviation widths of ($\sigma_{\mu_{\mathrm{\varphi_{1}}}},\sigma_{\mu_{\mathrm{\varphi_{2}}}})=(0.853,1.018)$ mas/yr, with an error on the mean of (0.174,0.208) mas/yr. In comparison the stream stars covering the same spatial range show ($\mu_{\mathrm{\varphi_{1}}},\mu_{\mathrm{\varphi_{2}}})=(0.064,0.051)$ mas/yr and widths ($\sigma_{\mu_{\mathrm{\varphi_{1}}}},\sigma_{\mu_{\mathrm{\varphi_{2}}}})=(0.557,0.792)$ mas/yr and an error on the mean of (0.055,0.079) mas/yr. In this case, the proper motions are significantly higher in the spur than in the accompanying stream, and even higher than the proper motion in the stream at the connecting point around $\varphi_{1}=-45$ degrees. This implies that the spur stars likely received a noticeable velocity kick away from the main stream due to the interaction with an external source. 

Finally, we consider the case of a gradient in proper motions across the spur region. To that end, we fit the data using a straight line ($f(x)=A(x+45) + B$) in the region $-45<\varphi_{1}<-20$ degrees, which is well populated. This results in solutions $(A,B)=(-0.010\pm0.011,0.121\pm0.164)$ for $\mu_{\mathrm{\varphi_{1}}}$ and $(A,B)=(0.010\pm0.010,0.090\pm0.165)$ for $\mu_{\mathrm{\varphi_{2}}}$. For comparison, we also fit the same line to the main stream, resulting in $(A,B)=(0.006\pm0.006,-0.167\pm0.088)$ for $\mu_{\mathrm{\varphi_{1}}}$ and $(A,B)=(-0.010\pm0.006,0.108\pm0.089)$ for $\mu_{\mathrm{\varphi_{2}}}$. The stream stars are consistent with a flat slope within 2 sigma in $\mu_{\mathrm{\varphi_{1}}}$, for both spur and stream stars. However, the proper motion gradient in $\mu_{\mathrm{\varphi_{2}}}$ is more pronounced and is pointing in the direction away from the stream for the spur stars. Considering Figure~\ref{GD1_resids} it becomes clear the number of stars populating the furthest part of the spur is small, meaning these gradient are subject to small number statistics and should be followed up with additional observations, most notably radial velocity measurements.

In the case of the blob feature, the PM residuals have ($\mu_{\mathrm{\varphi_{1}}},\mu_{\mathrm{\varphi_{2}}})=(0.058,0.067)$ mas/yr with widths ($\sigma_{\mu_{\mathrm{\varphi_{1}}}},\sigma_{\mu_{\mathrm{\varphi_{2}}}})=(0.892,0.769)$ mas/yr and the associated stream stars at similar $\varphi_{1}$ have ($\mu_{\mathrm{\varphi_{1}}},\mu_{\mathrm{\varphi_{2}}})=(0.184,0.002)$ mas/yr with width ($\sigma_{\mu_{\mathrm{\varphi_{1}}}},\sigma_{\mu_{\mathrm{\varphi_{2}}}})=(0.993,0.735)$ mas/yr. While slightly larger, the PM differences are still not significant enough (the error on the mean is (0.078,0.067) mas/yr for the blob and (0.161,0.119) mas/yr for the stream) to conclude they follow different orbits.

\section{Discussion} \label{sec:discussion}

\subsection{Alignment of GD-1}

Recently, \cite{carlberg_2019} argued that in $\Lambda$CDM, tidal streams should have velocities misaligned with the stream orientation due to late-time accretion events and the subsequent damped oscillations in the host galaxy's dark matter halo. They predict that the velocity perpendicular to the stream will in general be non-zero and that it should follow an exponential distribution with a scale velocity of $\sim10-20$ km/s. Using the technique suggested in \cite{orphan_lmc_mass}, we can compare the orientation of GD-1 on the sky with its proper motions to determine whether stars move along the stream. Similarly, we can compare the distance gradient along the stream with its radial velocity. 

\begin{figure*}
\centering
\includegraphics[angle=0, width=0.95\textwidth]{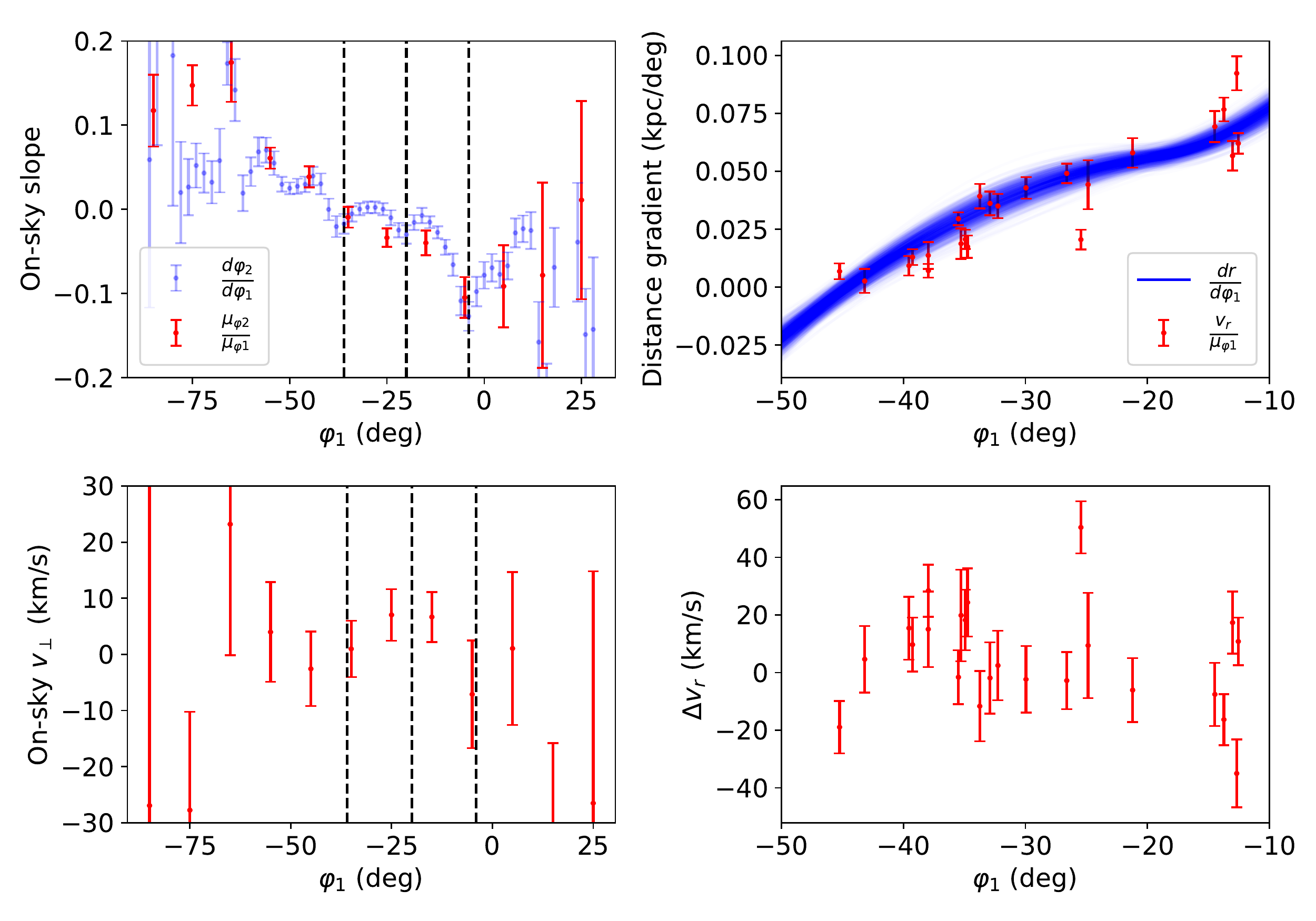}
\caption{Alignment of the GD-1 streams's kinematics and 3d structure. \textbf{Top left:} Comparison of the slope of the stream on the sky, $\frac{d\varphi_2}{d\varphi_1}$ (blue error bars), with the ratio of its reflex corrected proper motions, $\frac{\mu_{\varphi 2}}{\mu_{\varphi 1}}$ (red error bars). These are well matched across the entire range. Interestingly, stream track shows some sharp features which correspond to the location of the gaps in density (shown with vertical black dashed lines). \textbf{Bottom left:} On-sky velocity perpendicular to the stream track. This shows that the velocity perpendicular to the stream track is small and consistent with zero, i.e. that the stars are moving along the stream on the sky. \textbf{Top right:} Comparison of the slope of the stream in distance, $\frac{dr}{d\varphi_1}$, with the slope corresponding to its reflex corrected radial velocity and proper motions, $\frac{v_r}{\mu_{\varphi 1}}.$  The close match shows that the stream's velocity in the radial direction is pointed along the stream. The radial velocities used in this figure from come \protect\cite{Koposov10}. \textbf{Bottom right:} Difference between the stream's radial velocity and that expected from its proper motion and distance gradient. This difference is consistent with zero showing that the stream does not have a measurable velocity perpendicular to its track. } \label{fig:vel_align}
\end{figure*}

We show this alignment in Figure \ref{fig:vel_align}. In the top left panel we compare the slope of the stream on the sky, ${\rm d}\varphi_2/{\rm d}\varphi_1$, with the ratio of reflex corrected proper motions, ${\mu_{\varphi 2}}/{\mu_{\varphi 1}}$, as was done in Fig. 1 of \cite{orphan_lmc_mass}. We stress that this $\mu_{\varphi 1}$ does not have the typical $\cos\varphi_2$ factor. The close alignment shows that the stream stars are moving along the stream on the sky. Interestingly, we also see that the stream slope shows some rapid changes, i.e. wiggles in the stream track, which are aligned with the underdensities in the stream (shown with vertical black dashed lines). These correlated signals between the density and stream track are expected from subhalo impacts \citep[e.g.][]{Erkal15b}. Furthermore, the proper motions also appear to show the same rapid changes (e.g. around -5 deg and -45 deg), however this is not statistically significant. Improved data with \textit{Gaia} DR3 will allow us to determine whether these associated proper motion signals are real. 

In the bottom left panel of Figure \ref{fig:vel_align} we present an alternative way of looking at the stream alignment on the sky. We define the perpendicular velocity, $v_\perp$, as the velocity perpendicular to the local stream track. This is defined as 

\begin{equation}
\displaystyle v_\perp = \frac{kr\left( -\mu^*_{\varphi1} \frac{{\rm d}\varphi_2}{{\rm d}\varphi_1}\frac{1}{\cos \varphi_2} + \mu_{\varphi2} \right)}{\Big(1 + \left(\frac{{\rm d}\varphi_2}{{\rm d}\varphi_1} \frac{1}{\cos \varphi_2} \right)^2 \Big)^{1/2}} ,
\end{equation}
where $k=4.74047$ km s$^{-1}$ kpc$^{-1}$ mas$^{-1}$ and $r$ is the distance to each star. These velocities are small and consistent with zero although we note that the error bars are still quite significant. 

As with the proper motion on the sky, we can also compare the radial velocity, proper motion, and distance gradient. If the stream is an orbit these will be related, namely ${\rm d}r/{\rm d}\varphi_1 = {v_r}/\mu_{\varphi1}$. We compare these two quantities in the top right panel of Figure \ref{fig:vel_align} and we see that they match over the range where we have radial velocities. This shows that the stream is aligned in the radial direction and along the line of sight. Finally, we can use this gradient to determine the excess radial velocity along the line of sight, i.e.

\begin{equation}
\Delta v_r = v_r - \frac{{\rm d}r}{{\rm d}\varphi_1} \mu_{\varphi 1}  .  
\end{equation}
This quantity is shown in the bottom right panel of Figure \ref{fig:vel_align} and shows a scatter consistent with zero although the error bars are quite large. Improved measurements of the proper motions with \textit{Gaia} DR3 will allow us to better measure these perpendicular velocities for GD-1 and other streams and thus measure the time-dependence of the Milky Way halo as suggested in \cite{carlberg_2019}.  

\subsection{Orbit fits} \label{sec:orbits}

In order to further highlight how perturbed GD-1 is, we perform orbit fits in the Milky Way potential from \cite{mcmillan_2017} using \textsc{galpot} \citep{dehnen_binney_1998}. We choose to restrict the fit to where GD-1 has a significant density, $-60 < \varphi_1 < 10$ deg and to fix the potential to that of \cite{mcmillan_2017}. We also choose to integrate the orbit forwards and backwards from $\varphi_1 = -45$ deg. While the potential is fixed, we vary the other 5 parameters describing the orbit: radial velocity, proper motions, distance, and $\varphi_2$. The orbit is integrated forwards and backwards for 150 Myr. Given the orbit, we compute a combined likelihood using all observables (track on sky, distance, radial velocity, and proper motions). We use \textsc{emcee} to explore the likelihood space with 100 walkers and 1000 steps. 

\begin{figure*}
\centering 
\includegraphics[width=8.7cm]{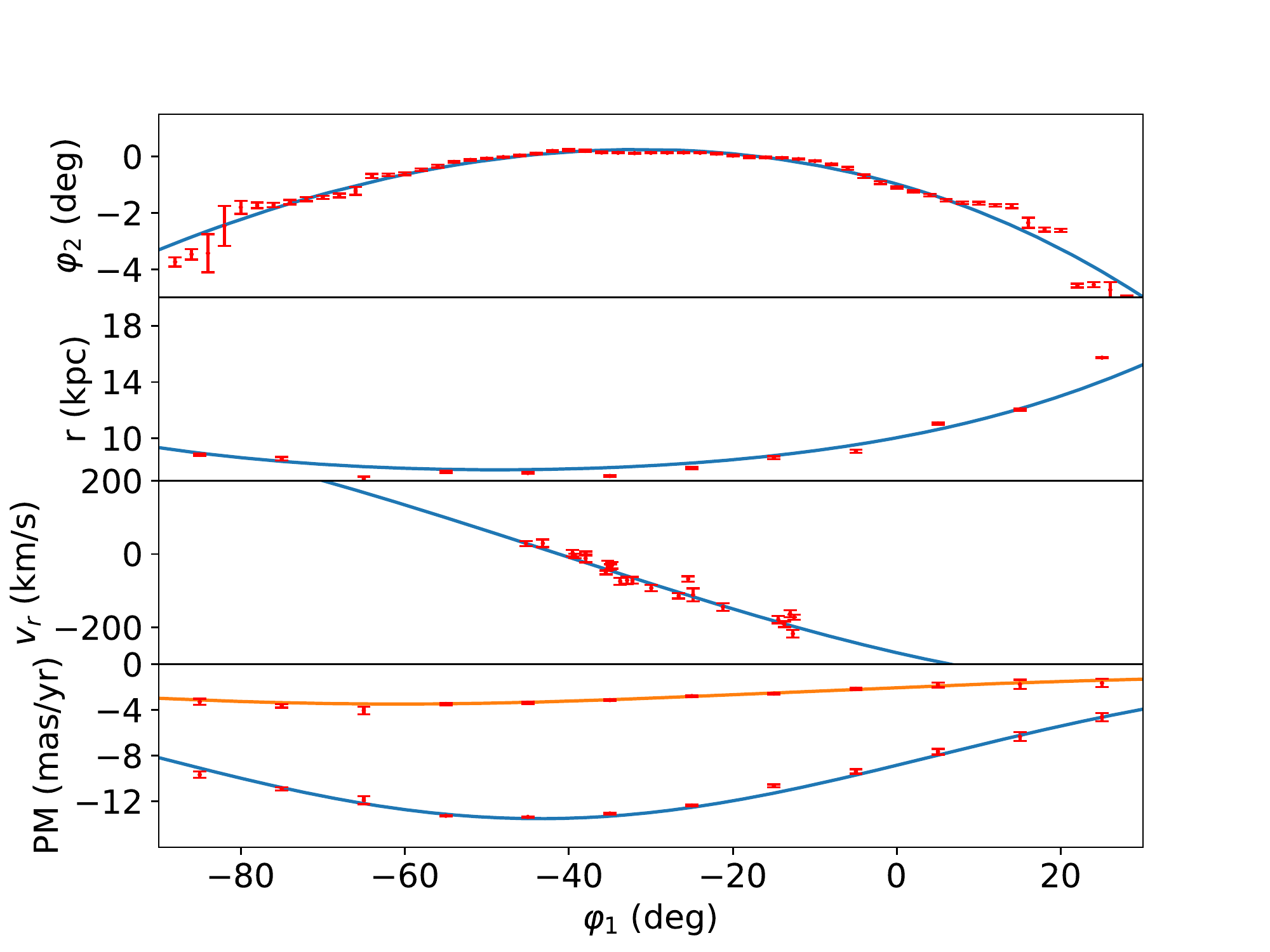}\quad
\includegraphics[width=8.7cm]{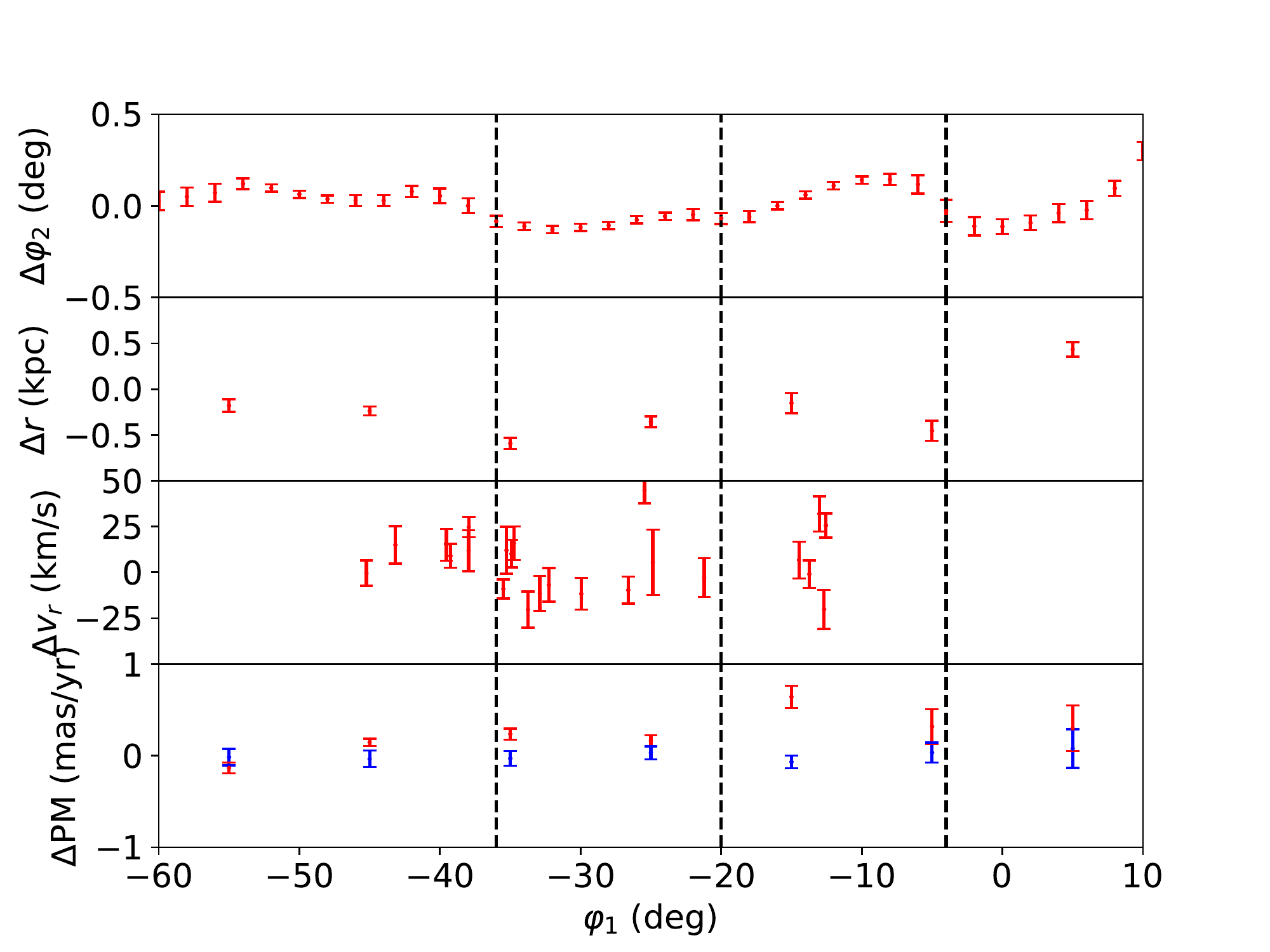}
\caption{Best fit orbit in the Milky Way potential from \protect\cite{mcmillan_2017}. The fit is restricted to the region where the stream has a significant density, i.e. $-60 < \varphi_1 < 10$ deg. The left panel shows the best-fit against the observables while the right panel shows the residuals for each observable. Interestingly, there are significant residuals in the stream track (top right panel) on small scales. These small scale residuals cannot be ameliorated by changing the Milky Way potential which will change the stream on larger scales. These residuals line up with the underdensities in the stream (vertical dashed-black lines) suggesting that they are connected. Such correlated signals are expected from perturbations by substructure \citep[e.g.][]{Erkal15b} }
\label{fig:orbit}
\end{figure*}

The best fit stream is shown in the left panel of Figure \ref{fig:orbit}. The stream matches all observables well, although the track deviates from the orbit outside of the range we fit. The right panel of Figure \ref{fig:orbit} shows the residuals of this best fit and the data. Interestingly, there are significant residuals in the stream track. These residuals are correlated with the underdensities in the stream (vertical dashed-black lines), especially the gap at $\varphi_1 = -4$ deg. These wiggles in the residual show that the stream track has significant small scale structure in it. This will not be ameliorated by changing the global Milky Way potential which will instead make large scale changes to the stream. 

These correlated signals are essential for determining the properties of the perturbers which created these features. Indeed, \cite{Erkal15b} showed that 3 observables are needed to fully fit the stream gap down to a 1d degeneracy between the subhalo's mass and relative velocity. For GD-1, there are now several features for which we have two observables (track on sky and density). Thus, radial velocity follow-up of these features is essential for completing the fits.

\subsection{Connection of the spur with other nearby structures}

\begin{figure}
\centering
\includegraphics[angle=0, width=0.495\textwidth]{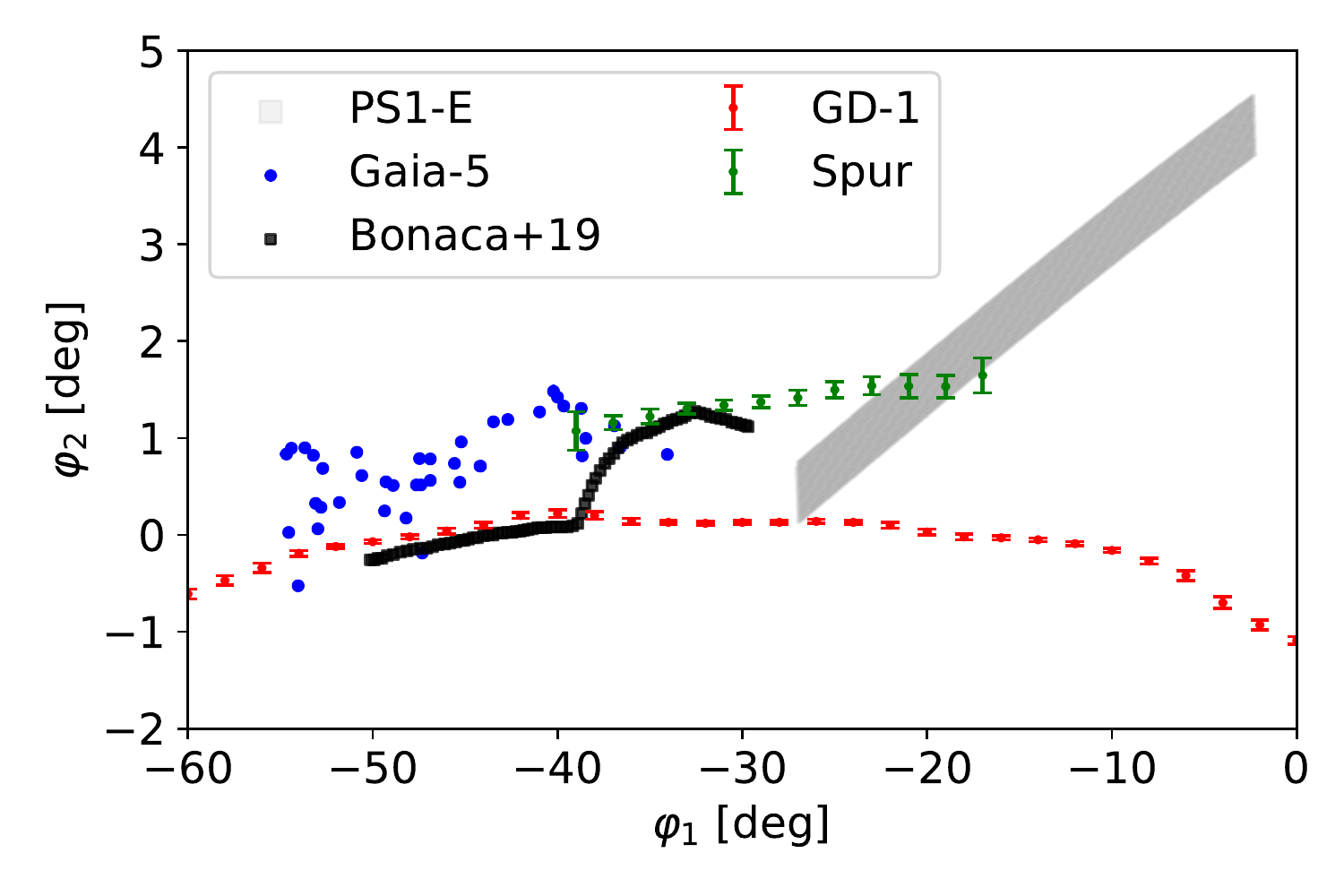}
\caption{Comparison of the results of this work with previously known features near GD-1. The black points show the spur from \protect\cite{Bonaca18}. The blue points show the stream, Gaia-5, found by \protect\cite{malhan_2018}. Although Gaia-5 looks like a continuation of the spur, it has a significantly different distance and proper motion than GD-1. The grey band shows the stream PS1-E discovered in \protect\cite{Bernard16}. Although this stream has a similar distance to GD-1, it has a different orientation to the spur when they overlap. Thus, although there appears to be a wealth of structure near GD-1, most of this is unrelated to the stream.  
} \label{fig:gd1_comp}
\end{figure}

In addition to the spur near GD-1 discovered by \cite{Price-Whelan18a}, there are a number of other stream-like features near GD-1. In particular there is the Gaia-5 stream discovered by \cite{malhan_2018} which is seemingly coincident with the spur. There is also the PS1-E stream discovered by \cite{Bernard16} which also looks like an extension of GD-1 \citep[see e.g. Fig. 1 of ][]{malhan_2019}. In Figure \ref{fig:gd1_comp} we show both of these streams in addition to our measured track for GD-1 and the spur. Note that in this figure, the track for PS1-E is taken from \textsc{galstreams} \citep{mateu_2018}. The points for Gaia-5 from \cite{malhan_2018} and for the spur as used in \cite{Bonaca18} are extracted from those papers. Although these streams are coincident on the sky, they are likely unrelated to GD-1 or the spur. Gaia-5 has a distance ($18.5$-$20.5$ kpc) which is much larger than GD-1 and a proper motion which is much smaller \citep{malhan_2018}. Similarly, although PS1-E has a distance of $\sim12.6$ kpc which is similar to GD-1 and the spur, the extended view of the spur in this work shows that the orientation of PS1-E is quite different in the region on the sky where they are co-located.

\subsection{Effect of the classical dwarfs on GD-1} \label{sec:Sgr}

Now we turn to the classical dwarfs to determine whether they can have an effect on GD-1. We use proper motions from \cite{gaiadr2_pms} and radial velocities and distances from \cite{McConnachie12}. We model each dwarf (except the LMC) as a Hernquist profile \citep{hernquist_1990} with a mass of $10^9 M_\odot$ and a scale radius of $1$ kpc. Since the LMC can affect the orbits of each satellite \citep{erkal_sat_orbit}, we must include the effect of the LMC.  We model the LMC as a Hernquist profile with a mass of $1.5\times10^{11} M_\odot$ and a scale radius of $17.14$ kpc, consistent with the measurement in \cite{orphan_lmc_mass}. We include the dynamical friction on the LMC from the Milky Way using the prescription of \cite{jethwa_2016}. We choose to model the Milky Way using the \texttt{MWPotential2014} from \cite{bovy_galpy} since it is computationally cheaper than the potential in \cite{mcmillan_2017}. Since we are using a different potential and since we include the LMC, we first re-fit GD-1 including the LMC. This cannot be done with orbits since different parts of the stream are affected differently by the LMC. 

Instead, we fit GD-1 using the stream generation technique from \cite{orphan_lmc_mass} which is based on the modified Lagrange Cloud Stripping technique from \cite{Gibbons14}. We chose to place the progenitor of GD-1 at $\varphi_1=-45^\circ$. Since there is no obvious progenitor over the observed extent of GD-1, we linearly interpolate the mass from an original value of $2\times10^{4} M_\odot$ 5 Gyr ago to zero at the present. We fit for its position on the sky, $\varphi_2$, distance, radial velocity, and proper motions. As in Section \ref{sec:orbits}, we define the likelihood using all of the observables and perform the fit using \textsc{emcee} \citep{emcee}. This gives a good fit for GD-1 and although we have included the LMC, these best-fits do not show any appreciable difference from the best-fit we show in Figure \ref{fig:orbit}.  

For each classical dwarf (except the LMC), we make 100 realizations of its present-day phase-space position. Using the best-fit stream parameters, the progenitor for GD-1 is then rewound for 5 Gyr in the combined presence of the dwarf, the LMC, and the Milky Way. Note that we account for the force of each of these components (i.e. GD-1 progenitor, dwarf, LMC, and Milky Way) on each other during this process. Thus, we also account for the reflex motion in the Milky Way from the LMC. The progenitor is then disrupted using the stream generation technique from \cite{orphan_lmc_mass} which is based on the modified Lagrange Cloud Stripping technique from \cite{Gibbons14}.

\begin{figure}
\centering
\includegraphics[angle=0, width=0.495\textwidth]{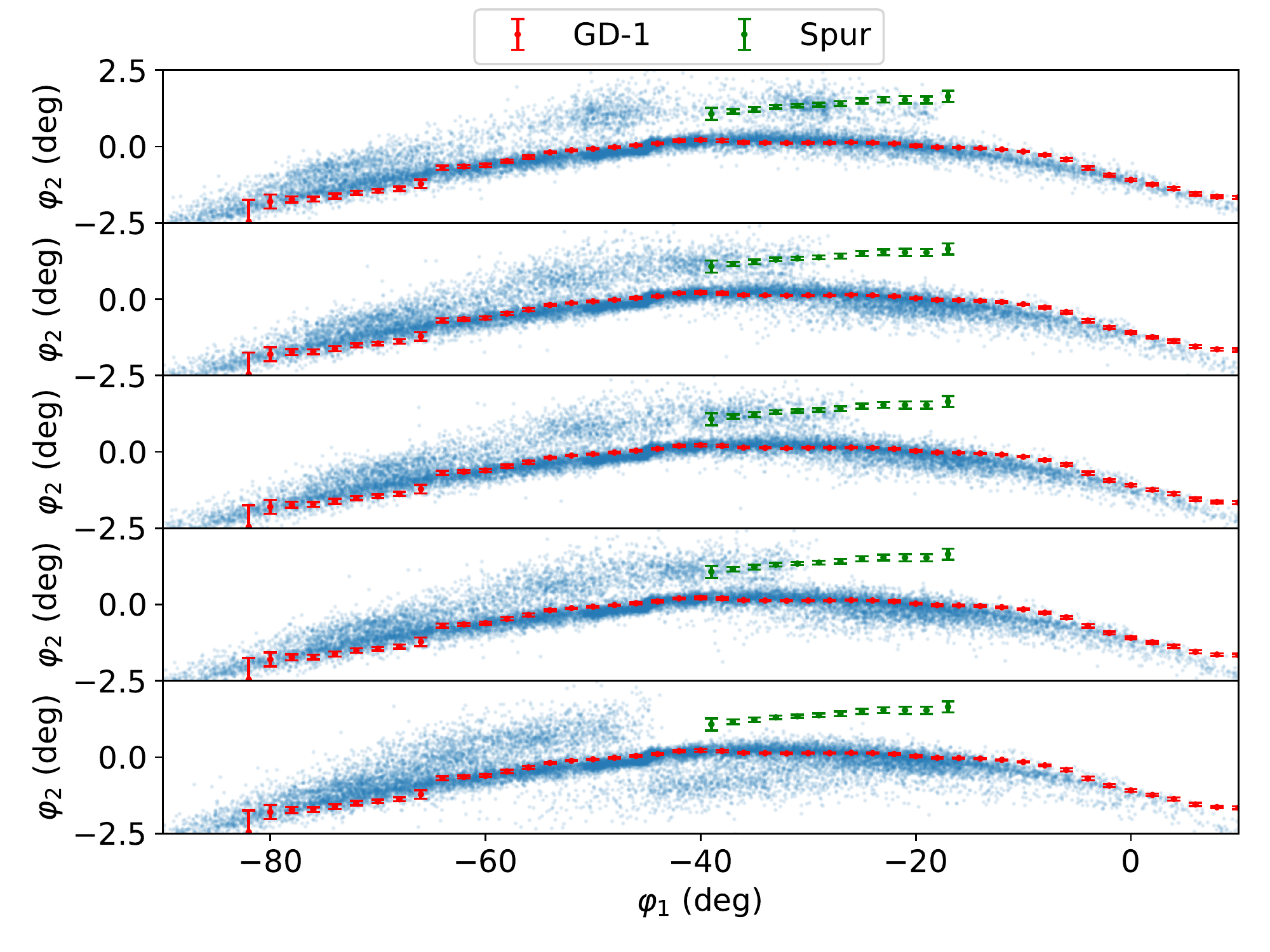}
\caption{Effect of the Sagittarius dwarf on the GD-1 stream. The panels show five different realizations of the present-day phase-space of Sagittarius. These were selected 
since they exhibited features similar to the spur. In each panel, the red error bars show the observed stream track and the green error bars show the observed track of the spur. In some of the realizations (i.e. lowest two panels) the interaction with Sagittarius can also create a feature below GD-1 in a similar location to where the blob.} \label{fig:gd1_sgr}
\end{figure}

Amongst the classical satellites, we find that only Sagittarius can have an appreciable effect on GD-1. This makes sense since the pericenter of Sagittarius is $\sim 15$ kpc \cite[e.g.][]{gaiadr2_pms} and it is the only classical dwarf which passes within the extent of GD-1's orbit. In order to study the effect of Sagittarius more closely, we perform an additional 900 realizations. A number of these realizations produce features similar to the spur in GD-1 and some also have features in the location of the blob. We show five of these realizations in Figure \ref{fig:gd1_sgr}. These features are due to a close encounter ($\sim $0.5-$3$ kpc) with Sagittarius approximately 3 Gyr ago. We stress that these five examples were chosen to have features like the spur and thus this is not a generic prediction of the effect of Sagittarius but rather a possibility given uncertainties on its present-day phase-space position. In order to assess whether the spur is due to the effect of Sagittarius or a substructure as modelled in \cite{Bonaca18}, we will need to compare the detailed morphology of the spur and better understand the past orbit of Sagittarius by getting a faithful fit to its stream. We note that we have ignored the disrupted dark matter halo of Sagittarius in this exploration and this could have a significant effect \citep[e.g.][]{bovy_2016}. 

Finally, we note that the perturbations from Sagittarius shown in Figure \ref{fig:gd1_sgr} feature a spur which extends much further to the left than the known spur. Interestingly, the observed width of GD-1 roughly doubles to the left of $\varphi_1 \sim -38^\circ$ which is close to where the observed spur terminates (see middle panel of Fig. \ref{GD1_stream_info}). Thus, it is possible that the spur is much longer than currently observed but is too close to GD-1 to distinguish it with current data. Deeper photometry in this region is needed to better understand how the spur connects onto GD-1 and how far it extends. 

\subsection{Morphology of the spur} \label{sec:spur_morphology}

In general, features like the spur, highlighted in Figure \ref{GD1_Gaia_trackzoom}, can naturally arise from a subhalo interaction. As a subhalo passes near the stream, it kicks stars and changes their orbits \citep[e.g.][]{Carlberg12,Erkal15}. The kicks along the stream direction change the orbital period of stars, causing them to subsequently move along the stream, while kicks perpendicular to the stream lead to oscillations about the stream. Generically, the perturbed streams will have wiggles in them. However, at late times, the perturbed stars overtake (or undertake) the unperturbed stars since they have different orbital periods and the stream folds on itself, giving the ``caustic" phase of gap growth identified in \cite{Erkal15}. In such a case, the stream folds on both the leading and trailing side of the gap. Indeed, \cite{Bonaca18} has provided the first fits to the spur where it arises from such a perturbation. 

\begin{figure}
\centering
\includegraphics[angle=0, width=0.495\textwidth]{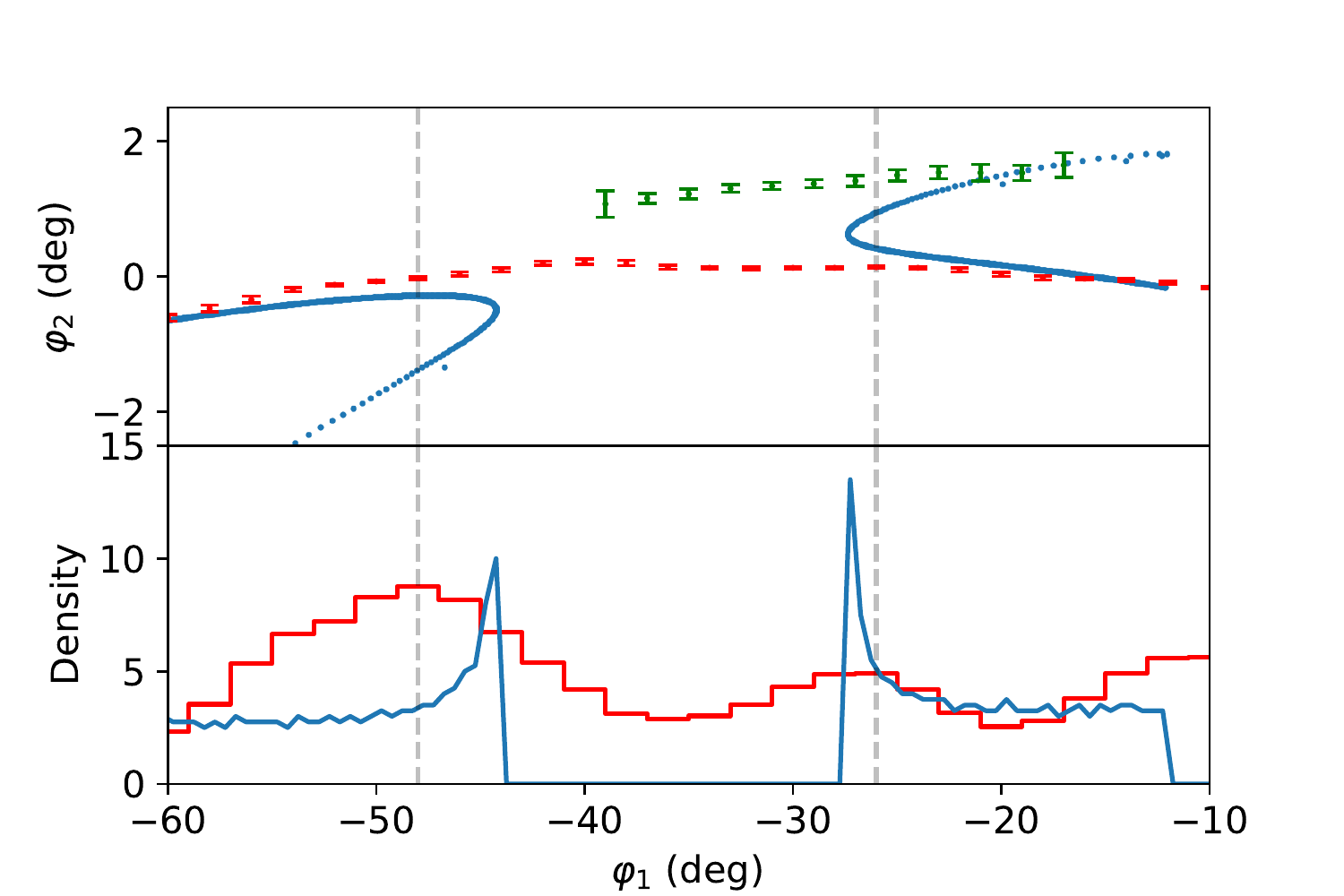}
\caption{Example of a spur created from a recent subhalo impact. The top panel shows the stream on the sky where blue points are the model and red and green error bars show the observed stream track and spur respectively. The bottom panel shows the stream density. The red histogram shows the observed density and the blue line shows the density of the model stream. The vertical black dashed lines show the location of the peaks in the stream density. Crucially, in the model the connection of the spur and the stream always lines up with the density peak in the stream. This is in contrast where the spur extends across most of the gap.} \label{fig:spur_ex}
\end{figure}

Interestingly, the extended view of the spur shown in this work is difficult to reconcile with the adjacent gap at $\varphi_1 \sim -36$ deg. This is because although features like the spur are generically expected, they also come with a precise relation between the stream track and the density. In particular, the peak in the density corresponds to the location where the spur attaches onto the stream. To demonstrate this point, in Figure \ref{fig:spur_ex} we show an example of a spur created from the impact of a stream with a subhalo. To do this, we start with the best-fit orbit from Section \ref{sec:orbits}, we then rewind the orbit for 1 Gyr and initialize a train of 1000 particles at 1 Gyr in the past using a spacing of 0.025 Myr. This train is then evolved to the present in the presence of the Milky Way potential and a subhalo modelled as a Hernquist profile. For the example shown in Figure \ref{fig:spur_ex}, we used a subhalo with a mass of $10^7 M_\odot$ and a scale radius of $0.25$ kpc. The closest approach with the stream (i.e. the particle train) occurs $~400$ Myr ago and gives rise to the spur feature. We note that this is not meant to be a fit to the stream but rather to highlight how the spur and density are connected. This also follows from the analytical results in \cite{Erkal15b}. This connection between the spur connection and density is also apparent in the models of \cite{Bonaca18} (e.g. their Fig. 3). 

Unlike the model, however, the observed spur extends across much of the gap (e.g. Fig. \ref{fig:spur_ex}). Thus, despite the close proximity of the spur and the gap, they may be unrelated. Indeed, \cite{Webb18} have recently suggested that the gap at $\sim -35^\circ$ may be due to the disruption of the progenitor. Instead, the spur could be explained by the model of Sagittarius we presented in Section \ref{sec:Sgr} or by a subhalo impact further to the left of the spur. \cite{carlberg_2018} has also argued that the original host dwarf galaxy which GD-1 was accreted with could significantly perturb the stream and \cite{malhan_2019} found debris around GD-1 which they argued was consistent with this picture.

\subsection{Mechanism for gap and wiggle at -3 deg} \label{sec:wiggle}

One other feature which stands out in the maps of GD-1 (e.g. see Fig. \ref{GD1_Gaia_trackzoom}) is the wiggle in the stream track and associated under-density at $\varphi\sim-3^\circ$. This feature was also found by \cite{malhan_2019}. They interpretted this feature as the location of the possible progenitor of GD-1. However, while the progenitor would be expected to be associated with a wiggle in the stream track \citep[see e.g. Fig. 12 in ][]{deBoer2018a}, this wiggle has the wrong orientation. The wiggle from a progenitor should be below and then above the average stream track as $\varphi_1$ decreases. This is the opposite to what is seen at $\varphi_1 \sim -3^\circ$. Thus, it appears that this wiggle and associated underdensity is likely a perturbation to the stream from substructure in the Milky Way. Such a feature is generically expected from a subhalo perturbation \citep[e.g.][]{Erkal15b}. Future fits to this feature will allow us to determine whether it is due to a baryonic substructure or a dark matter subhalo. 

\subsection{Association with globular clusters}

Using the orbits from Section \ref{sec:orbits}, we now compare the energy and actions of GD-1 with the population of GCs to look for possible associations.  These quantities are computed in the Milky Way potential from \cite{mcmillan_2017} using \textsc{agama} \citep{agama}. The globular cluster properties come from \cite{vasiliev_GCs}. To get the differences in energy and actions, we sample the chains from our fit to GD-1 in Section \ref{sec:orbits} and sample the observables for each GC from their errors. For the difference in actions, we add the differences of each action in quadrature, i.e. $\Delta J = \sqrt{\Delta J_z^2+\Delta J_\phi^2+\Delta J_r^2}$. 

\begin{figure}
\centering
\includegraphics[angle=0, width=0.495\textwidth]{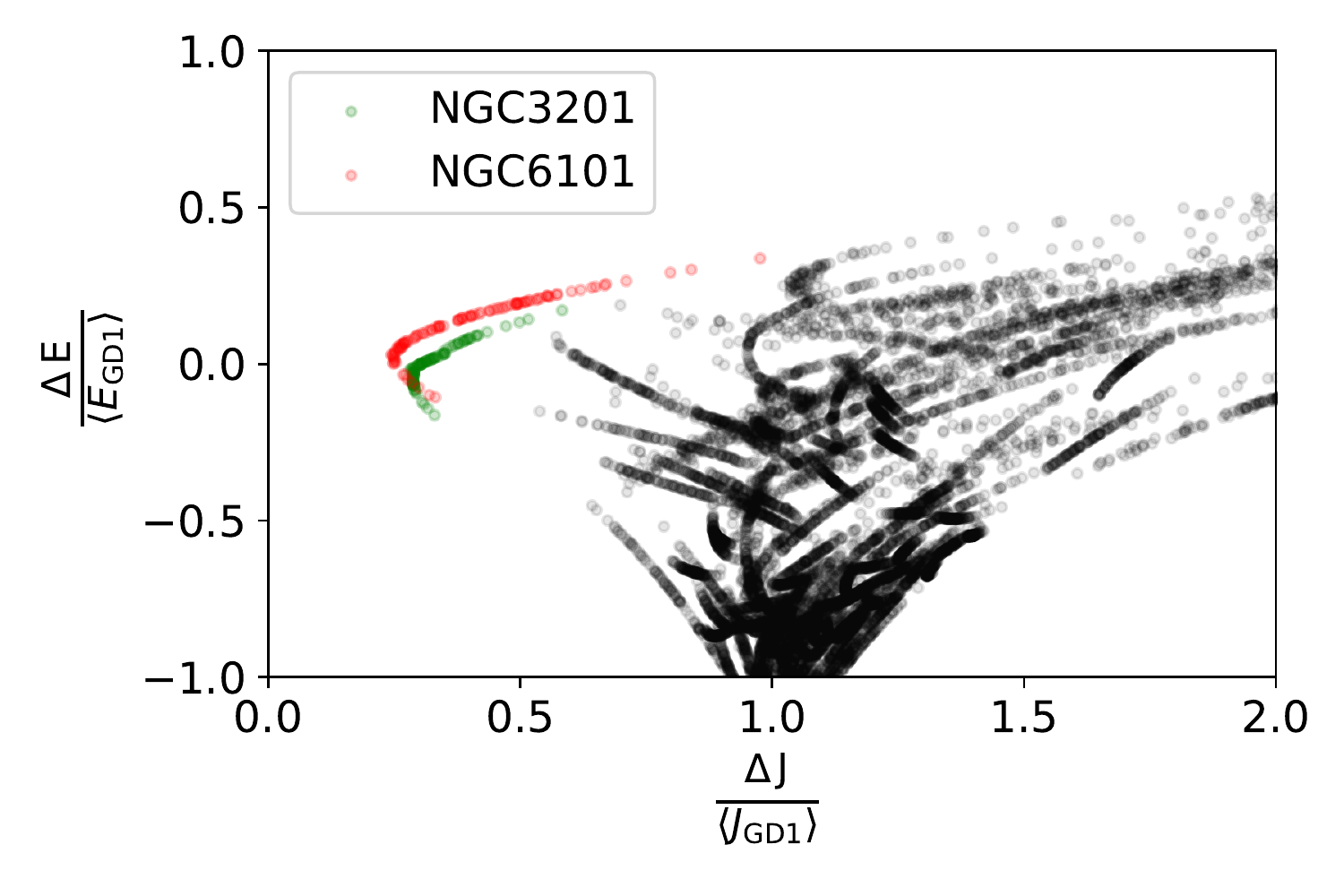}
\caption{Difference in energy and actions between GD-1 and the Milky Way globular clusters. Note that we have normalized the energy differences and action differences by the mean action and mean energy for our fits to GD-1. NGC 6101 and NGC 3201 both stand out as having the closest association in energy and action space.  Both of these globular clusters have been associated with the Sequoia merger \protect\citep{seqouia_myeong}, suggesting that the remnants of this merger will be on similar orbits to GD-1 and can thus have a substantial effect on the stream.} \label{fig:actions}
\end{figure}

The difference in energy and action is shown in Figure \ref{fig:actions} which shows that both NGC 3201 and NGC 6101 are the closest to GD-1 in phase space. Although neither of these is on a similar enough orbit to be the progenitor of GD-1, this is particularly interesting since both of these are believed to have been accreted with the Gaia-Sequoia event \citep{seqouia_myeong}. We note that \cite{seqouia_myeong} also point out the similarity in action between the Gaia-Sequoia clusters and GD-1. This suggests that GD-1 could be exposed to a higher background of dark matter subhaloes than a stream on a random orbit since the subhaloes of the Gaia-Sequoia galaxy will be on similar orbits to GD-1. Furthermore, these subhaloes will have smaller relative velocities to GD-1, allowing them to impart larger velocity kicks and thus perturb GD-1 more than naively expected. 

\section{Conclusions}\label{conclusions}
In this work, we have utilised data from {\it Gaia} DR2 to study the distance, morphology and density of stars in the GD-1 stream. Similar to other works, we stress that the precise PMs delivered by the {\it Gaia} mission were instrumental in selecting a high probability sample of members, which can hopefully be reproduced in other streams. 

Our analysis of the distance to GD-1 in Section~\ref{distance} results in values consistent with previous works across the often limited spatial range sampled~\citep{Koposov10, deBoer2018a}. However, the increased confidence of stream member selection using PMs from {\it Gaia} DR2 uncovers that the distance does not monotonically decrease across the leading tail (negative $\varphi_{1}$), but instead shows a minimum of $\approx$7 kpc around $\varphi_{1}\approx-50$ degrees followed by an increasing distance to $\approx$15 kpc. This has consequences for the possible orbits inhabited by GD-1 and will serve as a valuable constraint on models reproducing the GD-1 morphology. 

Furthermore, the increasing distance on both ends of the sampled stream footprint implies that GD-1 might still extend out further than currently determined. At current, stars at these distances do not have accurate PMs from the {\it Gaia} satellite, nor will they have good photometry in the relatively shallow wide-fill surveys such as SDSS and 2MASS. The most straightforward path to uncovering the continuation of the stream is by obtaining even deeper photometric data on either end, in order to probe the faint main sequence of the stream population. This will allow us to obtain a robust sample of high likelihood member stars and uncover the true extent of GD-1.

The PMs of GD-1 were studied in Section~\ref{proper_motions}, and probe to be instrumental in decoupling the stream motion from that of the MW foreground. Figure~\ref{GD1_PMfit} shows that the GD-1 stars stand out clearly in $\mu_{\mathrm{\varphi_{1}}}$ across most of the stream angle probed here. Separation in $\mu_{\mathrm{\varphi_{2}}}$ is less pronounced due to the relatively small sample size above g=20, but the stream over density is visible and consistent models of GD-1 (Erkal et al., in prep). We stress that the stars belonging to the spur and blob features show PMs very close to those of the main stream, and are therefore decoupled from the MW at high significance. The likelihood of these stars being drawn from the contaminating MW populations are very low, and any chance overlap with another retrograde  stream is unlikely. Therefore, we conclude that these features are genuinely associated to the GD-1 stream and must be taken into account in a full modelling of the stream.

The stream track of GD-1 was studied in Section~\ref{stream_track} after applying a Boolean matched filter procedure taking into account the distance and the PM cuts. The recovered track in Figure~\ref{GD1_Gaia_trackfit} shows small scale wiggles around the large scale stream track, along with several under dense gaps. The gaps and wiggles largely line up with recent works by \citet{deBoer2018a,Price-Whelan18a}  and confirm that GD-1 a long, narrow stream with significant disturbances. Notably, we uncover a striking sinusoidal wiggle in the stream track at high $\varphi_{1}$ straddling a gap feature at $\varphi_{1}\approx-3$ degrees. While this wiggle looks similar to characteristic ``S"-shape of stars coming off the stream's progenitor \citep[as in Pal 5,][]{pal5_disc}, this wiggle has the wrong orientation given the orbit of GD-1 \citep[e.g. Fig. 12 of][]{deBoer2018a}. Thus, this feature must instead come from a perturbation to the stream. Since GD-1 is on a retrograde orbit, the effect of baryonic substructure in the disk should be minimal \citep{Amorisco16} and thus this is a promising candidate for a subhalo interaction. 

The mass sampling corrected 1D density histograms of GD-1 (see Figure~\ref{GD1_stream_info}) show clear density variations across the $\approx$100 degrees of stream sampled here. Three significant under densities are visible at $\varphi_{1}=-36,-20$ and -3 deg, along with over dense peaks in between. This shows that GD-1 is likely to have experienced multiple disturbances and/or stripping episodes during the formation of its stream. Unfortunately, the location of the progenitor still remains a mystery, despite several promising candidate locations \citep[e.g.,][]{Webb18}. The total initial stellar mass of GD-1 is constrained from the matched filter procedure to be 1.58$\pm$0.07$\times$10$^{4}$ M$_{\odot}$, which is in line with expectations for a globular cluster origin. The stream mass is not unusually low compared to others \citep{Shipp18}, but GD-1 is much longer and narrowed than other streams of comparable mass. 

The track and PM stream residuals from extracted stream members in Figure~\ref{GD1_resids} show that the stars spatially belonging to the blob and spur are not significantly different from those of the main stream at similar $\varphi_{1}$. Given the absolute PMs at these locations, this makes it highly likely that both features are associated to GD-1 and must be taken into account when fully modelling the stream formation. Although small offsets in PM space are present, the PM accuracy and small sample size prevent us from seeing if the features are moving on different orbits than the main stream, at this point.

With the data in hand, we then fit the stream with an orbit and found a good match. Most interestingly, the stream has significant residuals in the stream track which line up with the location of the gaps in the density. These correlated signals will be crucial for fitting the perturbations to the stream, both by fitting each feature as suggested by \cite{Erkal15b} and for fitting the perturbations statistically while taking these correlations into account as suggested by \cite{Bovy17}. Equipped with this data set, we also showed that the stars in GD-1 are moving along the stream and do not have a significant velocity perpendicular to the stream. This is in contrast to the Orphan stream which shows a significant offset due to the effect of the LMC \citep{orphan_lmc_mass,kopsov_2019}. 

Finally, we showed that an interaction with Sagittarius could create features similar to the spur. In these models, the spur should be significantly longer than currently detected so this can be tested with future, deeper photometric data which shows how the spur connects onto the stream. We also argued that the spur is unlikely to be connected to the gap which is next to it since the observed spur extends across much of the gap instead of lining up with the edge of the gap as models predict.

The continuing investigation of the GD-1 stream reveals that it is highly complex with large and small scale variations and associated tidal debris. All of these will provide vital clues to the origin and evolution of the stream, which is a puzzle waiting to be unraveled.

\section*{Acknowledgements}
T.d.B. and M.G. acknowledge support from the European Research Council (ERC StG-335936). The authors also thank the International Space Science Institute (ISSI, Bern, CH) for welcoming the activities of Team 407 ``Globular Clusters in the Gaia Era". The authors also thank Jo Bovy and Jeremy Webb for useful comments on an early version of the manuscript, as well as Adrian Price-Whelan and Vasily Belokurov for useful discussions.

This work presents results from the European Space Agency (ESA) space mission {\it Gaia}. {\it Gaia} data are being processed by the Gaia Data Processing and Analysis Consortium (DPAC). Funding for the DPAC is provided by national institutions, in particular the institutions participating in the Gaia MultiLateral Agreement (MLA). The {\it Gaia} mission website is \url{https://www.cosmos.esa.int/gaia}. The Gaia archive website is \url{https://archives.esac.esa.int/gaia}.

This paper made used of the Whole Sky Database (wsdb) created by Sergey Koposov and maintained at the Institute of Astronomy, Cambridge by Sergey Koposov, Vasily Belokurov and Wyn Evans with financial support from the Science \& Technology Facilities Council (STFC) and the European Research Council (ERC).

The Pan-STARRS1 Surveys (PS1) and the PS1 public science archive have been made possible through contributions by the Institute for Astronomy, the University of Hawaii, the Pan-STARRS Project Office, the Max-Planck Society and its participating institutes, the Max Planck Institute for Astronomy, Heidelberg and the Max Planck Institute for Extraterrestrial Physics, Garching, The Johns Hopkins University, Durham University, the University of Edinburgh, the Queen's University Belfast, the Harvard-Smithsonian Center for Astrophysics, the Las Cumbres Observatory Global Telescope Network Incorporated, the National Central University of Taiwan, the Space Telescope Science Institute, the National Aeronautics and Space Administration under Grant No. NNX08AR22G issued through the Planetary Science Division of the NASA Science Mission Directorate, the National Science Foundation Grant No. AST-1238877, the University of Maryland, Eotvos Lorand University (ELTE), the Los Alamos National Laboratory, and the Gordon and Betty Moore Foundation.

\bibliographystyle{mn2e_fixed}
\bibliography{Bibliography}

\label{lastpage}

\end{document}